\documentclass[aps,prd,groupedaddress,showpacs,showkeys]{revtex4}
\usepackage{latexsym,epsfig,amsmath,amssymb}
\usepackage{amsfonts}
\usepackage{verbatim}
\newcommand{\eq}{\begin{eqnarray}}
\newcommand{\en}{\end{eqnarray}}

\begin{document}

\title{The scalar mesons in multi-channel $\pi\pi$ scattering
 and decays of the $\psi$ and $\Upsilon$ families}

\author{Yurii S.~Surovtsev$^1$, Petr Byd\v{z}ovsk\'y$^2$, Thomas Gutsche$^3$,
Robert~Kami\'nski$^4$, Valery~E.~Lyubovitskij$^{3,5}$,
Miroslav~Nagy$^6$ \vspace*{1.2\baselineskip}} \affiliation{
$^1$ Bogoliubov Laboratory of Theoretical Physics, Joint Institute for Nuclear
Research, Dubna 141980, Russia\\
\vspace*{.4\baselineskip}\\
$^2$ Nuclear Physics Institute, Czech Academy of Sciences, \v{R}e\v{z} near
Prague 25068, Czech Republic\\
\vspace*{.4\baselineskip}\\
$^3$ Institut f\"ur Theoretische Physik,  Universit\"at T\"ubingen,
Kepler Center for Astro and Particle Physics,
Auf der Morgenstelle 14, D--72076 T\"ubingen, Germany\\
\vspace*{.4\baselineskip}\\
$^4$ Institute of Nuclear Physics, Polish Academy of Sciences, Cracow 31342, Poland\\
\vspace*{.4\baselineskip}\\
$^5$ Department of Physics, Tomsk State University,
634050 Tomsk, Russia\\
\vspace*{.4\baselineskip}\\
$^6$ Institute of Physics, Slovak Academy of Sciences, Bratislava 84511, Slovak Republic}

\date{\today}

\begin{abstract}

The $f_0$ mesons are studied in a combined analysis of data on the isoscalar S-wave processes $\pi\pi\to\pi\pi,K\overline{K},\eta\eta$ and on decays $J/\psi\to\phi(\pi\pi, K\overline{K})$, $\psi(2S)\to J/\psi(\pi\pi)$, and $\Upsilon(2S)\to\Upsilon(1S)\pi\pi$ from the Argus, Crystal Ball, CLEO, CUSB, DM2, Mark~II, Mark~III, and BES~II collaborations. The method of analysis, based on
analyticity and unitarity and using an uniformization procedure, is set forth with
some details. Some spectroscopic implications from results of the analysis are discussed.

\end{abstract}



\maketitle

\section{Introduction}

The problem of scalar mesons, particularly their nature, parameters,
and status of some of them, is still not solved~\cite{PDG12}. E.g., applying our method of the uniformizing variable in the 3-channel analyses of multi-channel
$\pi\pi$ scattering \cite{SBKN-PRD10,SBL-PRD12} we have obtained parameters of
the $f_0(500)$ and $f_0(1500)$ which differ considerably from results of analyses
utilizing other methods (mainly based on dispersion relations or Breit-Wigner
approaches).
Reasons for this difference were understood in our works  \cite{SBKLN-PRD12,SBKLN-1206_3438,SBKLN-1207_6937}. We have shown that when studying
wide multi-channel resonances, as the scalar ones, the Riemann-surface structure
of the $S$-matrix of considered processes must be allowed for properly. For the
scalar states this is, as minimum, the 8-sheeted Riemann surface. This is related
with a necessity to analyze jointly coupled processes
$\pi\pi\to\pi\pi,K\overline{K},\eta\eta$ because, as it was shown, studying only
$\pi\pi$ scattering it is impossible to obtain correct values for parameters of
the scalar states. Calculating masses, total widths and coupling constants of
resonances with channels, one must use the poles on sheets II, IV and VIII,
depending on the state type.

From these results an important conclusion can be drawn: Even if a wide resonance
does not decay into a channel which opens above its mass but it is strongly connected
with this channel, one ought to consider this state taking into account the
Riemann-surface sheets related to the threshold branch-point of this channel.
I.e., the dispersion relation approach in which amplitudes are considered
only on the 2-sheeted Riemann surface does not suit for correct determination of this
resonance parameters.

Note importance of our above-indicated conclusions because our approach is based
only on the demand for analyticity and unitarity of the amplitude using an
uniformization procedure. The construction of the amplitude is essentially free
from any dynamical (model) assumptions utilizing only the {\it mathematical} fact
that a local behaviour of analytic functions determined on the Riemann surface is
governed by the nearest singularities on all corresponding sheets. Therefore it
seems that our approach permits us to omit theoretical prejudice in extracting
the resonance parameters.

Analyzing only $\pi\pi\to\pi\pi,K\overline{K},\eta\eta(\eta\eta^\prime)$ \cite{SBL-PRD12}
in the 3-channel approach, we have shown that experimental data on the $\pi\pi$
scattering below 1 GeV admit two possibilities for parameters of the $f_0(500)$
with mass, relatively near to the $\rho$-meson mass, and with the total widths
about 600 and 950~MeV -- solutions ``A'' and ``B'', respectively.

Furthermore, it was shown that for the states $f_0(1370)$, $f_0(1500)$ (as
a superposition of two states, broad and narrow), and $f_0(1710)$, there are
four scenarios of possible representation by poles and zeros on the Riemann
surface giving similar descriptions of the above processes and, however,
quite different parameters of some resonances. E.g., for the $f_0(500)$ (A solution),
$f_0(1370)$ and $f_0(1710)$, a following spread of values is obtained for
the masses and total widths respectively: 605-735 and 567-686~MeV, 1326-1404
and 223-345~MeV, and 1751-1759 and 118-207~MeV.

Adding to the combined analysis the data on decays
$J/\psi\to\phi(\pi\pi, K\overline{K})$ from the Mark~III, DM2 and BES~II
collaborations \cite{Mark_III}, we have considerably diminished a quantity
of the possible scenarios \cite{SBKLN-1207_6937}. Moreover {\it the di-pion
mass distribution in the $J/\psi\to\phi\pi\pi$ decay of the BES~II data from
the threshold to about 850~MeV prefers surely the solution with the wider
$f_0(500)$ -- B-solution.} This is a problem because most of physicists \cite{PDG12}
prefer a narrower $f_0(500)$. Therefore, here we expanded our combined
analysis adding also accessible data on the decays $\psi(2S)\to J/\psi(\pi\pi)$
and $\Upsilon(2S)\to\Upsilon(1S)\pi\pi$ from the Argus, Crystal Ball, CLEO, CUSB,
and Mark~II collaborations \cite{Mark_II,Argus}.

There are also other problems related to interpretation of scalar mesons,
e.g., as to an assignment of the scalar mesons to lower $q{\bar q}$ nonets.
There is a number of properties of the scalar mesons, which do not allow to
make up satisfactorily the lowest nonet. The main of them is inaccordance of
the approximately equal masses of the $f_0(980)$ and $a_0(980)$ and the found
$s{\bar s}$ dominance in the wave function of the $f_0(980)$. If these states are
in the same nonet, the $f_0(980)$ must be heavier by 250-300 MeV than $a_0(980)$
because the difference of masses of $s$- and $u$-quarks is 120-150 MeV.
In connection with this, various variants for solution are proposed. The most
popular one is the 4-quark interpretation of ${f_0}(980)$ and ${a_0}(980)$ mesons,
in favour of which as though additional arguments have been found on the basis of
interpretation of the experimental data on the decays
$\phi\to\gamma\pi^0\pi^0,\gamma\pi^0\eta$ \cite{Achasov-4q}. However, the 4-quark
model, beautifully solving the old problem of the unusual properties of scalar mesons,
sets new questions. Where are the 2-quark states, their radial excitations and the
other members of 4-quark multiplets $9,9^*,36$ and $36^*$, which are predicted to
exist below 2.5 GeV \cite{Jaffe-4q}? We proposed our way to solve this problem.

Existence of the $f_0(1370)$ meson is still not obvious. In some works, e.g.,
in \cite{MO-02-2,Ochs10} one did not find any evidence for the existence of the
$f_0(1370)$. On the other hand, in Ref.~\cite{Bugg1370} a number of data requiring
apparently the existence of the $f_0(1370)$ is indicated. We have shown \cite{SBL-PRD12}
that an existence of the $f_0(1370)$ does not contradict the data on processes
$\pi\pi\to\pi\pi,K\overline{K},\eta\eta(\eta\eta^\prime)$ and if this state exists,
it has a dominant $s{\bar s}$ component. In the hidden gauge unitary approach,
the $f_0(1370)$ appears dynamically generated as a $\rho\rho$
state~\cite{Geng_Oset} and the $f_0(1710)$ as generated from the
$K^\ast \bar K^\ast$ interaction.

Further we shall consider mainly the 3-channel case because it was shown that this
is a minimal number of channels needed for obtaining correct values of scalar resonance
parameters. However for convenience and having in mind other problems, we shall mention
sometimes the 2- and N-channel cases.

\section{Method of the uniformizing variable in the 3-channel $\pi\pi$ scattering}

\begin{figure}[!thb]
\begin{center}
\includegraphics[width=0.5\textwidth,angle=0]{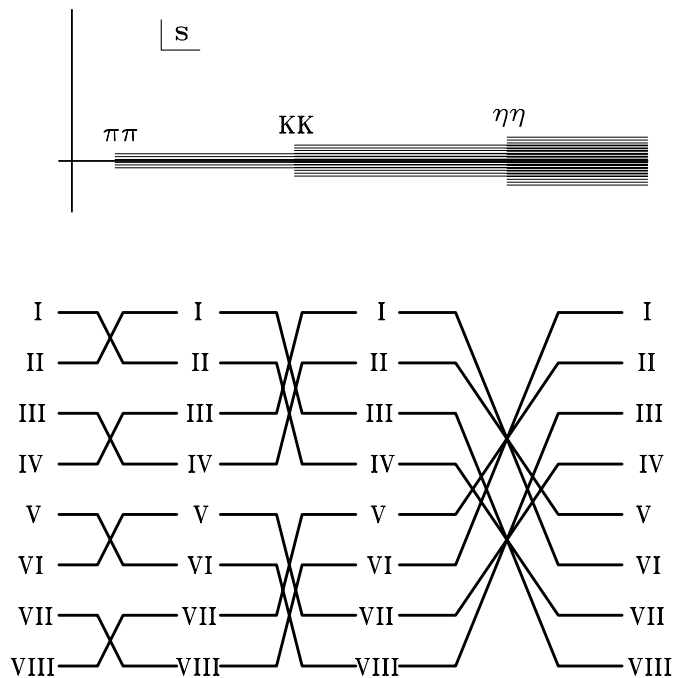}
\vspace*{-0.cm}\caption{Sewing together the sheets of the Riemann surface. \label{Fig:Sheets_sewing}}
\end{center}
\end{figure}

Our model-independent method which essentially utilizes a uniformizing variable can be
used only for the 2-channel case and under some conditions for the 3-channel one. Only
in these cases we obtain a simple symmetric (easily interpreted) picture of the resonance
poles and zeros of the $S$-matrix on the uniformization plane. The 2- or 3-channel
$S$-matrix is determined on the 4- or 8-sheeted Riemann surface, respectively. The matrix
elements $S_{ij}$, where $i,j=1,2,3$ denote channels, have the right-hand cuts along the
real axis of the $s$ complex plane ($s$ is the invariant total energy squared), starting
with the channel thresholds $s_i$ ($i=1,2,3$), and the left-hand cuts related to the
crossed channels. The Riemann-surface sheets, denoted by the Roman numbers, are numbered according to the signs of analytic continuations of the square roots $\sqrt{s-s_i}$ as
follows
\begin{center}
\def\arraystretch{1.2}
\begin{tabular}{|c|cccccccc|} \hline
{} & ~~I~~ & ~~II~~ & ~III~ & ~IV~ & ~~V~~ & ~VI~ & ~VII~ & ~VIII~ \\ \hline
{$\mbox{Im}\sqrt{s-s_1}$} & $+$ & $-$ & $-$ & $+$ & $+$ & $-$ & $-$ & $+$ \\
{$\mbox{Im}\sqrt{s-s_2}$} & $+$ & $+$ & $-$ & $-$ & $-$ & $-$ & $+$ & $+$\\
{$\mbox{Im}\sqrt{s-s_3}$} & $+$ & $+$ & $+$ & $+$ & $-$ & $-$ & $-$ & $-$\\
\hline
\end{tabular}
\end{center}
The sewing together of the Riemann surface sheets is shown in Fig.~\ref{Fig:Sheets_sewing}.

Our approach is based on general principles, as analyticity and unitarity, and
realizes an idea of the consistent account of the nearest (to the physical region)
singularities on all sheets of the Riemann surface of the $S$-matrix, thus giving
a chance to obtain a model-independent information on multi-channel resonances from
the analysis of data on the coupled processes. The main model-independent contribution
of resonances is given by poles and corresponding zeros on the Riemann surface.
A reasonable and simple description of the background should be a criterion of
correctness of this statement. Obviously, we deal with renormalized quantities,
and the poles of S-matrix correspond to dressed particles.

If a resonance has the only decay mode (1-channel case), the general statement about
a behaviour of the process amplitude is that at energy values in a proximity to the
resonance the amplitude describes the propagation of resonance as if it is a free
particle. This means that in the matrix element the resonance (in the limit of its
narrow width) is represented by a pair of complex conjugate poles on sheet II and by
a pair of conjugate zeros on the physical sheet at the same points of complex energy.
This model-independent statement about the poles as the nearest singularities holds
also when taking account of the finite width of a resonance.
Obviously, the statement that the poles corresponding to resonances are the nearest
(to the physical region) singularities holds also in the multi-channel case.

In order to obtain an arrangement of poles and zeros of multi-channel resonance
on the Riemann surface, we use the proved fact that on the physical sheet,
the $S$-matrix elements can possess only the resonance zeros (beyond the real axis),
at least, around the physical region. Therefore it is necessary to obtain formulas
expressing analytic continuations of the $S$-matrix elements to all sheets in terms
of those on the physical sheet \cite{KMS-nc96}.

To this end, let us consider the $N$-channel $S$-matrix (all channels are two-particle
ones) determined on the $2^N$-sheeted Riemann surface. The Riemann surface has the
right-hand (unitary) cuts along the real axis of the $s$-variable complex plane
$(s_i,\infty)$~ ($i=1,2,\cdots,N$ means a channel) through which the physical sheet
is sewed together with other sheets. The branch points are at the vanishing values of
the channel momenta~ ${k_\alpha}=(s/4-m_\alpha^2)^{1/2}$. For now we will neglect
the left-hand cut in the Riemann-surface structure related with the crossing-channel contributions, whose contribution, in principle, can be taken into account in the
background of the corresponding amplitudes.

In the following it is convenient to use enumeration of sheets (see, e.g., \cite{Kato-AP65}):
the physical sheet is denoted as $L_0$, other sheets through $L_{i_1\cdots i_k}$ where ~$i_1\cdots i_k$~ are a system of subscripts of those channel-momenta $k_{i_n}$ which change signs
at analytical continuations from the physical sheet onto the indicated one.
Then the analytical continuations of $S$-matrix elements $S_{ik}$ to the unphysical sheet
$L_{i_1\cdots i_k}$ are $S_{ik}^{(i_1\cdots i_k)}$. We will obtain the formula expressing
$S_{ik}^{(i_1\cdots i_k)}$ in terms of $S_{ik}^{(0)}$ (matrix elements $S_{ik}$ on the
physical sheet $L_0$), using the reality property of the analytic functions and the
$N$-channel unitarity. The direct derivation of these formulas requires rather bulky
algebra. It can be simplified if we use Hermiticity of the $K$-matrix.

To this end, first, we shall introduce the notation:
~${\bf S}^{[i_1\cdots i_k]}$~ means a matrix in which all the rows are
composed of the vanishing elements but the rows $i_1,\cdots,i_k,$ that consist of elements
$S_{i_n i_m}$. In the matrix ~${\bf S}^{\{i_1\cdots i_k\}}$, on the contrary, the rows
$i_1,\cdots,i_k$ are zeros. Therefore,
\begin{equation} \label{}
{\bf S}^{[i_1\cdots i_k]}+{\bf S}^{\{i_1\cdots i_k\}}={\bf S}.
\end{equation}
Further ~${\bf \Delta}^{[i_1\cdots i_k]}$~ and ~${\bf \Delta}^{\{i_1\cdots i_k\}}$~ denote
the diagonal matrices with the elements
\begin{displaymath}
\Delta_{ii}^{[i_1\cdots i_k]}=
\left\{\begin{array}{l}
1\quad\mbox{if}~~i\in (i_1\cdots i_k),\\
0\quad\mbox{for remaining}~~i, \end{array}\right.\quad\mbox{and} \quad
\Delta_{ii}^{\{i_1\cdots i_k\}}=\left\{\begin{array}{l}
0\quad\mbox{if}~~i\in (i_1\cdots i_k),\\
1\quad\mbox{for remaining}~~i, \end{array}\right.
\end{displaymath}
respectively. Further using relation of the $S$- and $K$-matrices
\begin{equation} \label{}
{\bf S} =\frac{I+i{\bf \rho}^{1/2}{\bf K}{\bf \rho}^{1/2}}
{I-i{\bf \rho}^{1/2}{\bf K}{\bf \rho}^{1/2}}~~~{\rm where}~~\rho_{ij}=0~(i\neq j),~~\rho_{ii}={2k_i}/{\sqrt{s}}
\end{equation}
and~ ${\bf S}{\bf S}^+={\bf I}$, it is easy to obtain that ${\bf K}={\bf K}^+$, i.e., the $K$-matrix has no discontinuity when going across the two-particle unitary cuts and has the same value in all sheets of the Riemann surface of the $S$-matrix. Using the latter fact, we obtain the needed formula. The analytical continuations of the $S$-matrix to the sheet $L_{i_1\cdots i_k}$ will be represented as
\begin{equation} \label{}
{\bf S}^{(i_1\cdots i_k)}=\frac{{\bf S}^{(0)\{i_1\cdots i_k\}}-
i{\bf \Delta}^{[i_1\cdots i_k]}}{{\bf \Delta}^{\{i_1\cdots i_k\}}-
i{\bf S}^{(0)[i_1\cdots i_k]}}.
\end{equation}
From the last formula the corresponding relations for the $S$-matrix elements can be derived by the formula for the matrix division. In Table~\ref{tab:An_contin} the result is shown for the 3-channel case. We have returned to more standard enumeration of sheets by Roman numerals I, II,...,VIII.
\begin{table}[!htb]
\begin{center}
{
\large
\def\arraystretch{1.5}
\begin{tabular}{|c|cccccccc|}\hline
{} & $L_{0}$ & $L_{1}$ & $L_{12}$ & $L_{2}$ & $L_{23}$ & $L_{123}$ & $L_{13}$ & $L_{3}$ \\
{Process} & I & II & III & IV & V & VI & VII & VIII \\
\hline $1\to 1$ & $S_{11}$ & $\frac{1}{S_{11}}$ &
$\frac{S_{22}}{D_{33}}$ & $\frac{D_{33}}{S_{22}}$ & $\frac{\det
S}{D_{11}}$
& $\frac{D_{11}}{\det S}$ & $\frac{S_{33}}{D_{22}}$ & $\frac{D_{22}}{S_{33}}$\\
$1\to 2$ & $S_{12}$ & $\frac{iS_{12}}{S_{11}}$ &
$\frac{-S_{12}}{D_{33}}$ & $\frac{iS_{12}}{S_{22}}$
& $\frac{iD_{12}}{D_{11}}$ & $\frac{-D_{12}}{\det S}$ & $\frac{iD_{12}}{D_{22}}$ & $\frac{D_{12}}{S_{33}}$\\
$2\to 2$ & $S_{22}$ & $\frac{D_{33}}{S_{11}}$ &
$\frac{S_{11}}{D_{33}}$ & $\frac{1}{S_{22}}$ &
$\frac{S_{33}}{D_{11}}$
& $\frac{D_{22}}{\det S}$ & $\frac{\det S}{D_{22}}$ & $\frac{D_{11}}{S_{33}}$\\
$1\to 3$ & $S_{13}$ & $\frac{iS_{13}}{S_{11}}$ &
$\frac{-iD_{13}}{D_{33}}$ & $\frac{-D_{13}}{S_{22}}$
& $\frac{-iD_{13}}{D_{11}}$ & $\frac{D_{13}}{\det S}$ & $\frac{-S_{13}}{D_{22}}$ & $\frac{iS_{13}}{S_{33}}$\\
$2\to 3$ & $S_{23}$ & $\frac{D_{23}}{S_{11}}$ &
$\frac{iD_{23}}{D_{33}}$ & $\frac{iS_{23}}{S_{22}}$
& $\frac{-S_{23}}{D_{11}}$ & $\frac{-D_{23}}{\det S}$ & $\frac{iD_{23}}{D_{22}}$ & $\frac{iS_{23}}{S_{33}}$\\
$3\to 3$ & $S_{33}$ & $\frac{D_{22}}{S_{11}}$ & $\frac{\det
S}{D_{33}}$ & $\frac{D_{11}}{S_{22}}$ & $\frac{S_{22}}{D_{11}}$ &
$\frac{D_{33}}{\det S}$ & $\frac{S_{11}}{D_{22}}$ &
$\frac{1}{S_{33}}$\\\hline
\end{tabular}}
\caption{Analytic continuations of the 3-channel $S$-matrix elements to unphysical sheets. \label{tab:An_contin}}
\end{center}
\end{table}

In Table~\ref{tab:An_contin}, the superscript $I$ is omitted to simplify the notation,
$\det S$ is the determinant of the $3\times3$ $S$-matrix on sheet~I, $D_{\alpha\beta}$ is the minor of the element $S_{\alpha\beta}$, that is, $D_{11}=S_{22}S_{33}-S_{23}^2$, $D_{22}=S_{11}S_{33}-S_{13}^2$, $D_{33}= S_{11}S_{22}-S_{12}^2$, $D_{12}=S_{12}S_{33}-S_{13}S_{23}$, $D_{23}= S_{11}S_{23}-S_{12}S_{13}$, etc.

These formulas show how singularities and resonance poles and zeros are transferred from the matrix element $S_{11}$ to matrix elements of coupled processes. Starting from the resonance zeros on sheet~I, one can obtain the arrangement of poles and zeros of resonance on the whole Riemann surface.

Let us explain in the 2-channel example how a pole cluster describing resonance arises.
In the 1-channel consideration of the scattering $1\to1$ the main model-independent
contribution of resonance is given by a pair of conjugate poles on sheet~II and by a pair
of conjugate zeros on sheet I at the same points of complex energy in $S_{11}$.
(\underline{Conjugate} poles and zeros are needed for real analyticity.)
In the 2-channel consideration of the processes $1\to1$, $1\to2$ and $2\to2$, we have
\begin{eqnarray}
&&S_{11}^{\rm II}=\frac{1}{S_{11}^{\rm I}},\qquad~~~~~~~~~~~~~
S_{11}^{\rm III}=\frac{S_{22}^{\rm I}}{S_{11}^{\rm
I}S_{22}^I-(S_{12}^{\rm I})^2}, \quad
S_{11}^{\rm IV}=\frac{S_{11}^{\rm I}S_{22}^{\rm I}
-(S_{12}^{\rm I})^2}{S_{22}^I},\\
&&S_{22}^{\rm II}=\frac{S_{11}^{\rm I}S_{22}^{\rm I}-(S_{12}^{\rm
I})^2}{S_{11}^{\rm I}},\quad S_{22}^{\rm III}=\frac{S_{11}^{\rm
I}}{S_{11}^{\rm I}S_{22}^{\rm I}-(S_{12}^{\rm I})^2},\quad
S_{22}^{\rm IV}=\frac{1}{S_{22}^{\rm I}},\\
&&S_{12}^{\rm II}=\frac{iS_{12}^{\rm I}}{S_{11}^{\rm
I}},\qquad~~~~~~~~~~~~~ S_{12}^{\rm III}=\frac{-S_{12}^{\rm
I}}{S_{11}^{\rm I}S_{22}^{\rm I}-(S_{12}^{\rm I})^2},\quad
S_{12}^{\rm IV}=\frac{iS_{12}^{\rm I}}{S_{22}^{\rm I}}.
\label{S_contin}
\end{eqnarray}
In $S_{11}$ a resonance is represented by a pair of conjugate poles on sheet~II and by a pair of conjugate zeros on sheet~I and also by a pair of conjugate poles on sheet~III and by a pair of conjugate zeros on sheet~IV at the same points of complex energy if the coupling of channels is absent ($S_{12}=0$). If the resonance decays into both channels and/or takes part in exchanges in the crossing channels, the coupling of channels arises ($S_{12}\not=0$). Then positions of the poles on sheet~III (and of corresponding zeros on sheet~IV) turn out to be shifted with respect to the positions of zeros on sheet~I.
Thus we obtain the cluster (of type ({\bf a})) of poles and zeros.

\underline{In the 2-channel case}, {\it 3 types} of resonances are obtained corresponding to a pair of conjugate zeros on sheet I only in $S_{11}$ -- the type ({\bf a}), only in $S_{22}$ -- ({\bf b}), and simultaneously in $S_{11}$ and $S_{22}$ -- ({\bf c}).\\
\underline{In the 3-channel case}, we obtain {\it 7 types} of resonances corresponding to 7 possible situations when there are resonance zeros on sheet I only in $S_{11}$ -- ({\bf a}); ~~$S_{22}$ -- ({\bf b}); ~~$S_{33}$ -- ({\bf c}); ~~$S_{11}$ and $S_{22}$ -- ({\bf d}); ~~$S_{22}$ and $S_{33}$ -- ({\bf e}); ~~$S_{11}$ and $S_{33}$ --
({\bf f}); ~~$S_{11}$, $S_{22}$ and $S_{33}$ -- ({\bf g}). The resonance of every type is represented by the pair of complex-conjugate \underline{clusters} (of poles and zeros on the Riemann surface).

{\it A necessary and sufficient condition for existence of the multi-channel resonance is its representation by one of the types of pole clusters.}  A main \underline{model-independent} contribution of resonances is given by the pole clusters and possible remaining small (\underline{model-dependent}) contributions of resonances can be included in the background. This is confirmed further by the obtained very simple description of the background.

The cluster type is related to the nature of state. {\it E.g.}, if we consider the $\pi\pi$, $K\overline{K}$ and $\eta\eta$ channels, then a resonance, coupled relatively more strongly to the $\pi\pi$ channel than to the $K\overline{K}$ and $\eta\eta$ ones is described by the cluster of type ({\bf a}). In the opposite case, it is represented by the cluster of type ({\bf e}) (say, the state with the dominant $s{\bar s}$ component). The glueball must be represented by the cluster of type ({\bf g}) as a necessary condition for the ideal case.
Whereas cases ({\bf a}), ({\bf b}) and ({\bf c}) can be related to the
resonance representation by Breit-Wigner forms, cases ({\bf d}), ({\bf e}), ({\bf f}) and ({\bf g}) practically are lost at the Breit-Wigner description.

One can formulate {\it a model-independent test as a necessary condition} to distinguish a bound state of colorless particles ({\it e.g.}, a $K\overline{K}$ molecule) and a $q{\bar q}$ bound state \cite{KMS-nc96,Morgan_Penn_93}.

In the 1-channel case, the existence of the particle bound-state means the presence of a pole on the real axis under the threshold on the physical sheet.

\underline{In the 2-channel case}, existence of the bound-state in channel~2 ($K\overline{K}$ molecule) that, however, can decay into channel~1 ($\pi\pi$ decay), would imply the presence of the pair of complex conjugate poles on sheet~II under the second-channel threshold without the corresponding shifted pair of poles on sheet III.

\underline{In the 3-channel case}, the bound state in channel~3 ($\eta\eta$) that, however, can decay into channels 1 ($\pi\pi$ decay) and 2 ($K\overline{K}$ decay), is represented by the pair of complex conjugate poles on sheet II and by the pair of shifted poles on sheet III under the $\eta\eta$ threshold without the corresponding poles on sheets VI and VII.

According to this test, earlier we rejected interpretation of the $f_0(980)$ as the $K\overline{K}$ molecule because this state is represented by the cluster of type ({\bf a}) in the 2-channel analysis of processes $\pi\pi\to\pi\pi,K\overline{K}$ and, therefore, does not satisfy the necessary condition to be the $K\overline{K}$ molecule \cite{KMS-nc96}.

It is convenient to use the Le Couteur-Newton relations \cite{LeCou}. They express the $S$-matrix elements of all coupled processes in terms of the Jost matrix determinant $d(k_1,\cdots,k_N)\equiv d(s)$ that is a real analytic function with the only branch-points at $k_i=0$:
\begin{equation} \label{}
S_{ii}(s)=\frac{d^{(i)}(s)}{d(s)},
\end{equation}
\begin{equation} \label{}
\left|{\begin{array}{ccc}
S_{i_1i_1}(s) & \cdots & S_{i_1i_k}(s) \\
\vdots & \vdots & \vdots \\
S_{i_ki_1}(s) & \cdots & S_{i_ki_k}(s)
\end{array}} \right|=\frac{d^{(i_1\cdots i_k)}(s)}{d(s)}.
\end{equation}
Rather simple derivation of these relations, using the $ND^{-1}$ representation of amplitudes and Hermiticity of the $K$-matrix, can be found in Ref. \cite{Kato-AP65}. The analytical structure of the $S$-matrix on all Riemann sheets given above is thus expressed in a compact way by these relations.
The real analyticity implies
\begin{equation} \label{}
d(s^*)=d^*(s)~~~~{\rm for~all}~ s.
\end{equation}
The unitarity condition requires further restrictions on the $d$-function for physical $s$-values which will be discussed below in the example of 3-channel $S$-matrix.

In order to use really the representation of resonances by various pole clusters, it ought to transform our multi-valued $S$-matrix, determined on the 8-sheeted Riemann surface, to one-valued function. But that function can be uniformized only on torus with the help of a simple mapping. This is unsatisfactory for our purpose. Therefore, we neglect the influence of the lowest ($\pi\pi$) threshold branch-point (however, unitarity on the $\pi\pi$ cut is taken into account). This approximation means the consideration of the nearest to the physical region semi-sheets of the Riemann surface of the $S$-matrix. In fact, we construct a 4-sheeted model of the initial 8-sheeted Riemann surface that is in accordance with our approach of a consistent account of the nearest singularities on all the relevant sheets.

\begin{figure}[htb]
\begin{center}
\includegraphics[width=0.4\textwidth,angle=-90]{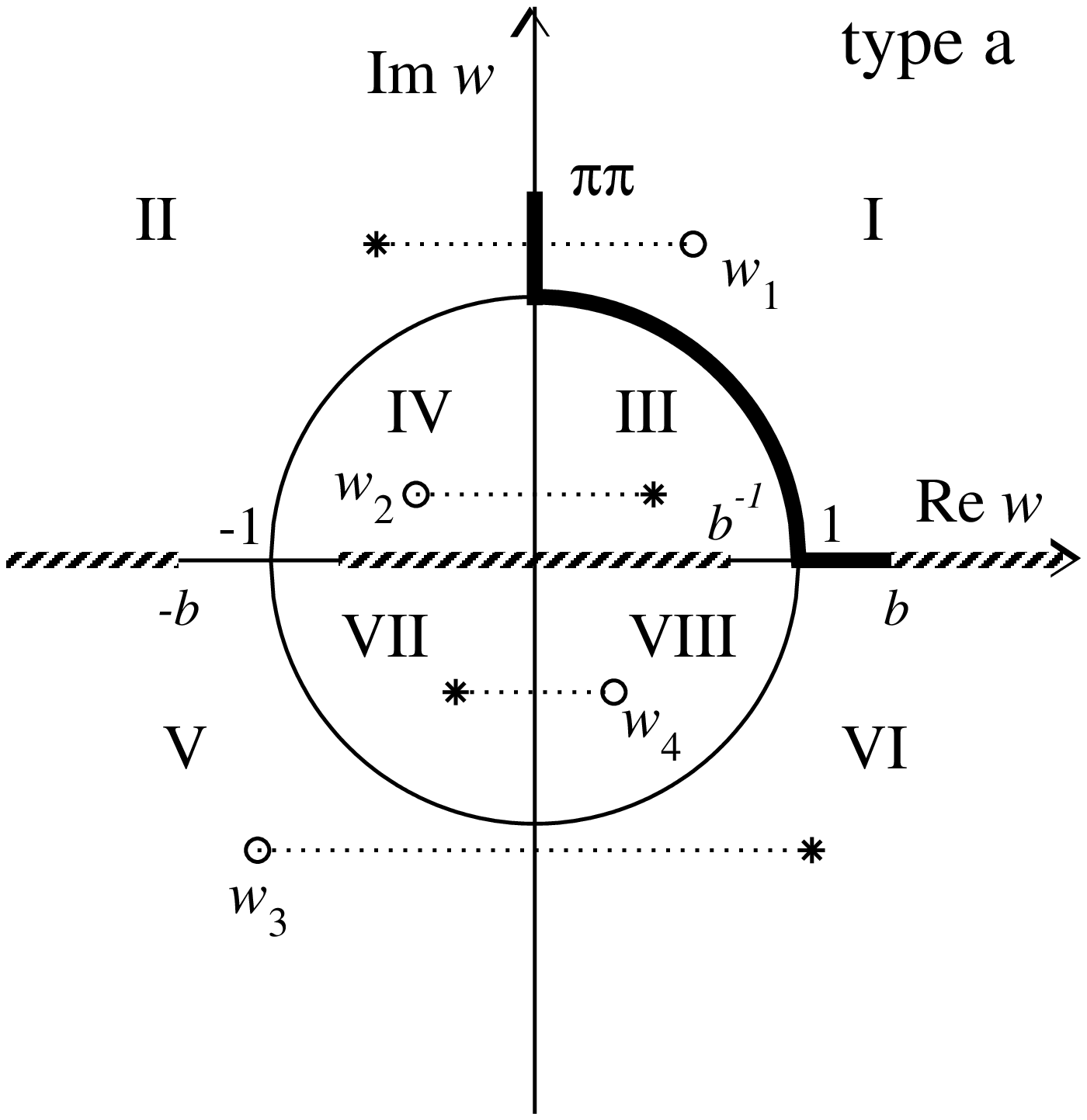}
\includegraphics[width=0.4\textwidth,angle=-90]{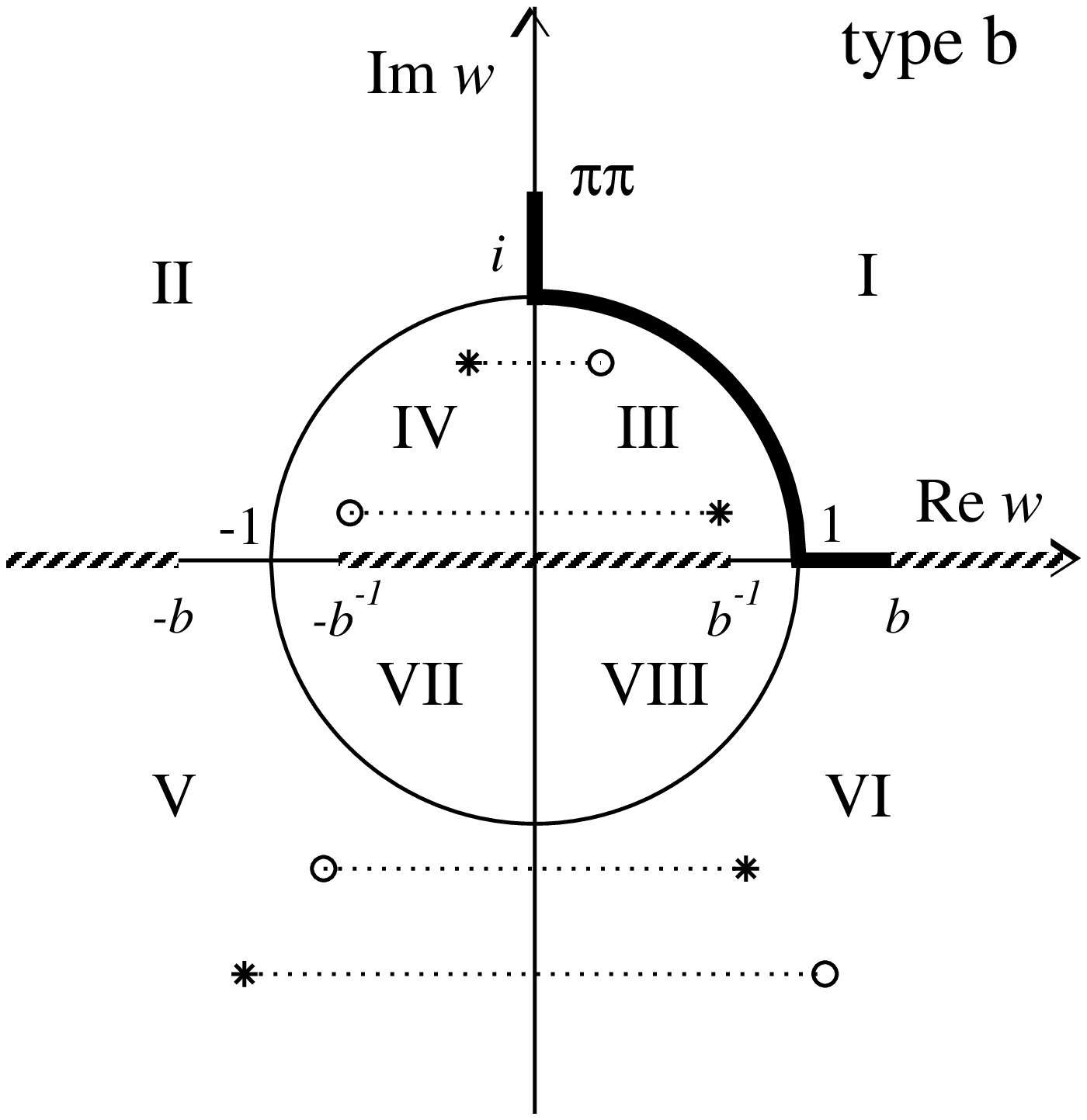}\\
\includegraphics[width=0.4\textwidth,angle=-90]{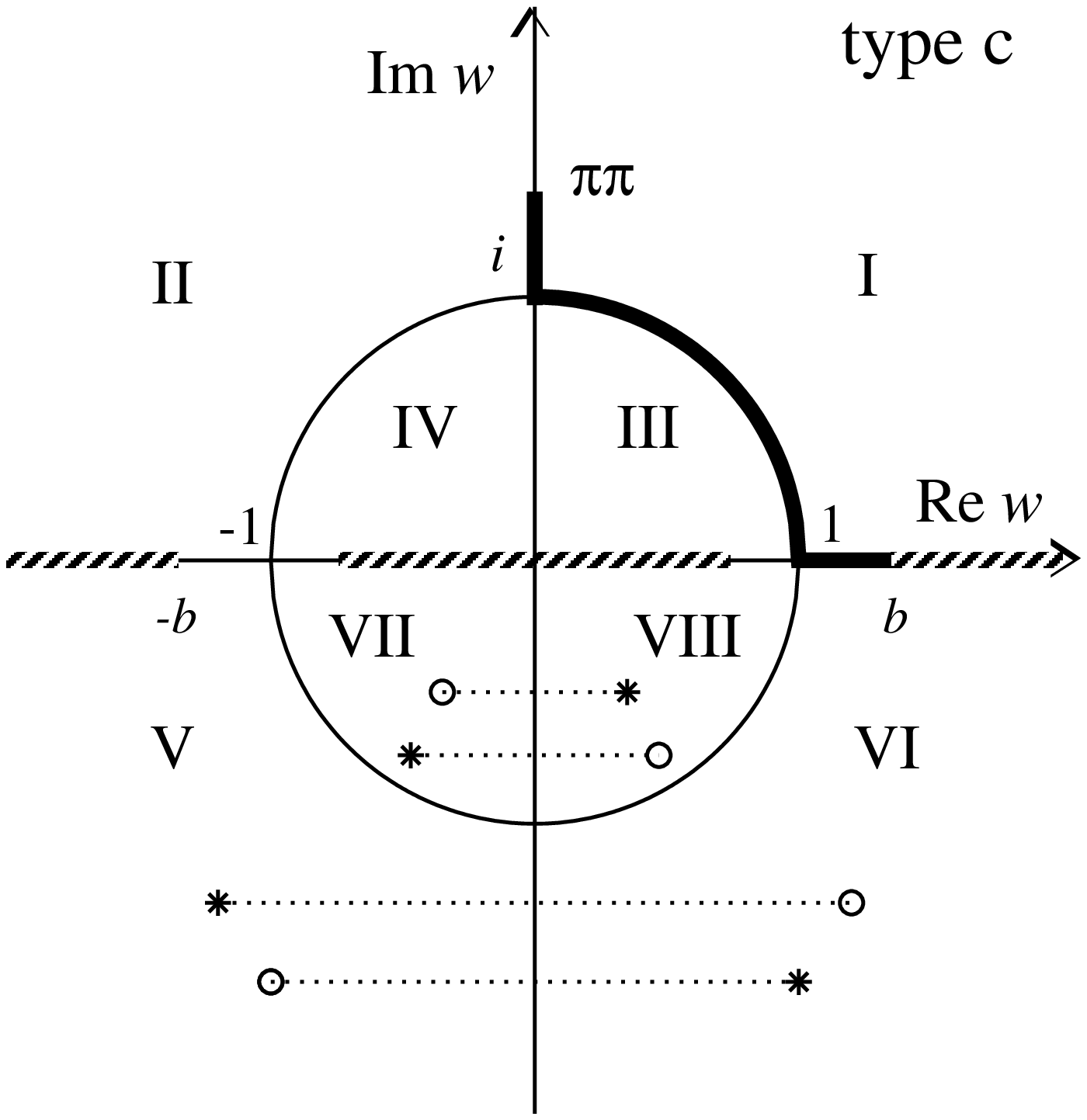}
\includegraphics[width=0.4\textwidth,angle=-90]{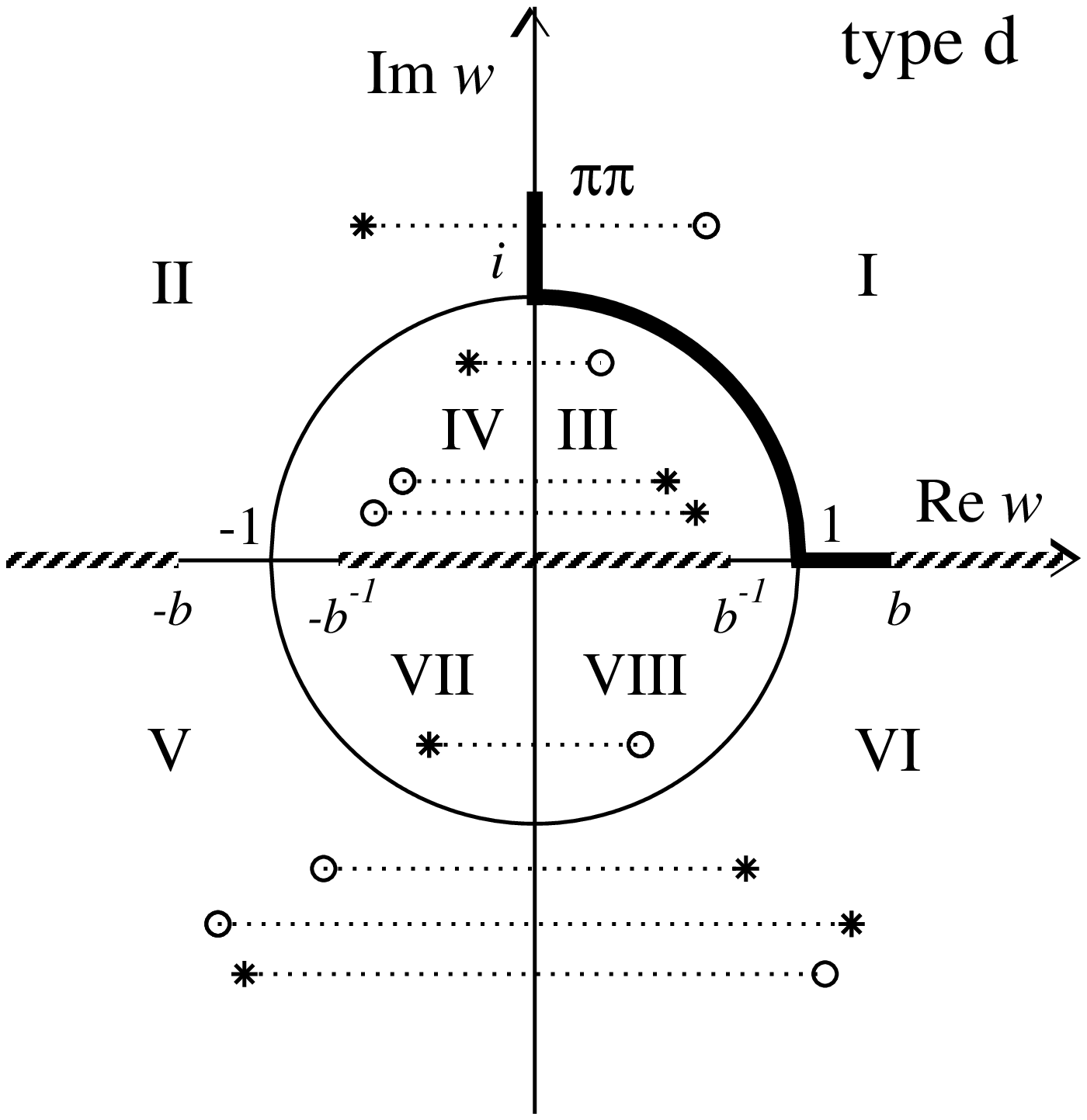}
\vspace*{-0.2cm}
\caption{Uniformization $w$-plane for the 3-channel-$\pi\pi$-scattering
matrix element. Representation of resonances of types ({\bf a}), ({\bf b}),
({\bf c}) and ({\bf d}) is shown. \label{fig:lw_plane_abcd}}
\end{center}
\end{figure}

\begin{figure}[htb]
\begin{center}
\includegraphics[width=0.4\textwidth,angle=-90]{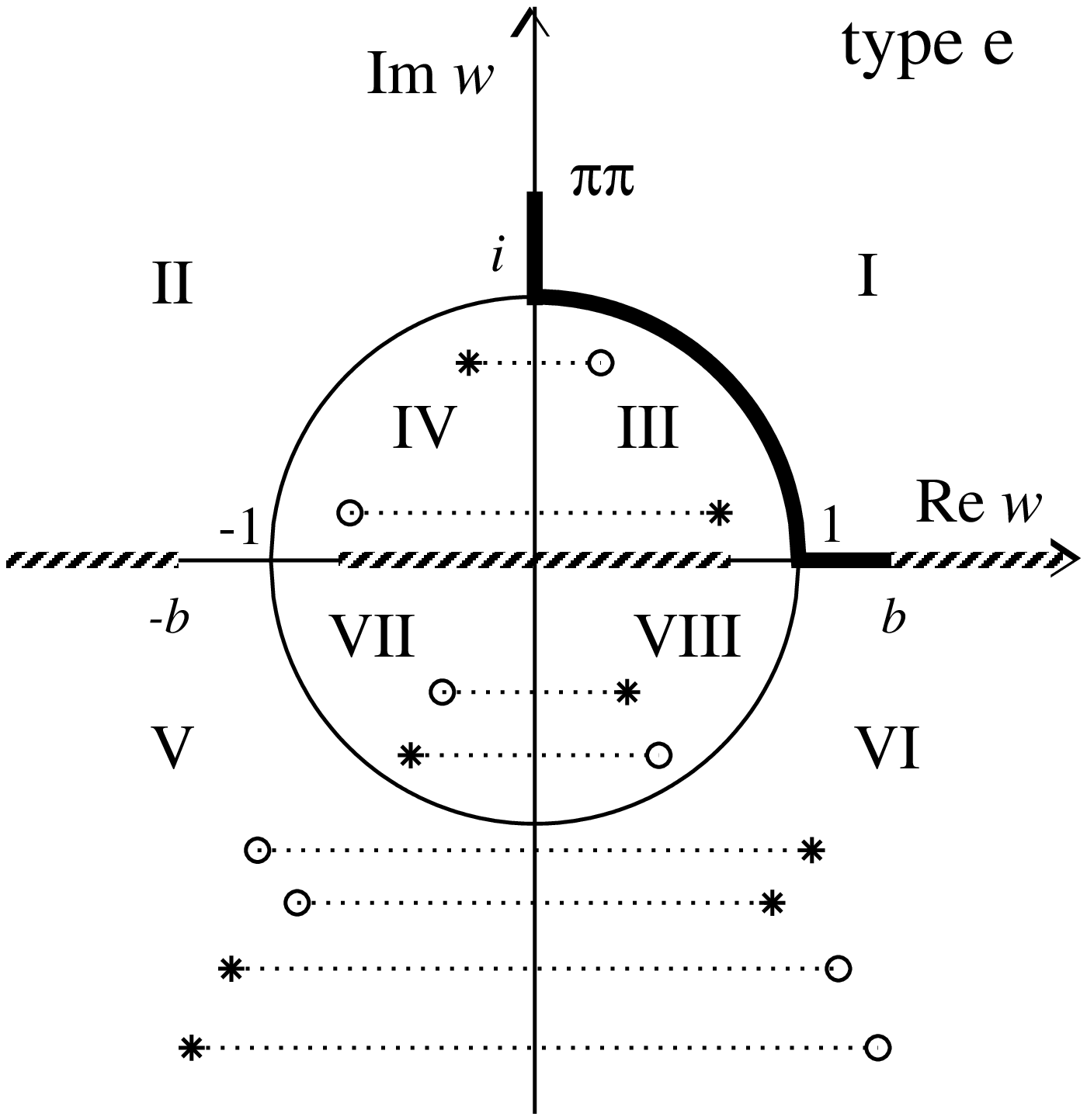}
\includegraphics[width=0.4\textwidth,angle=-90]{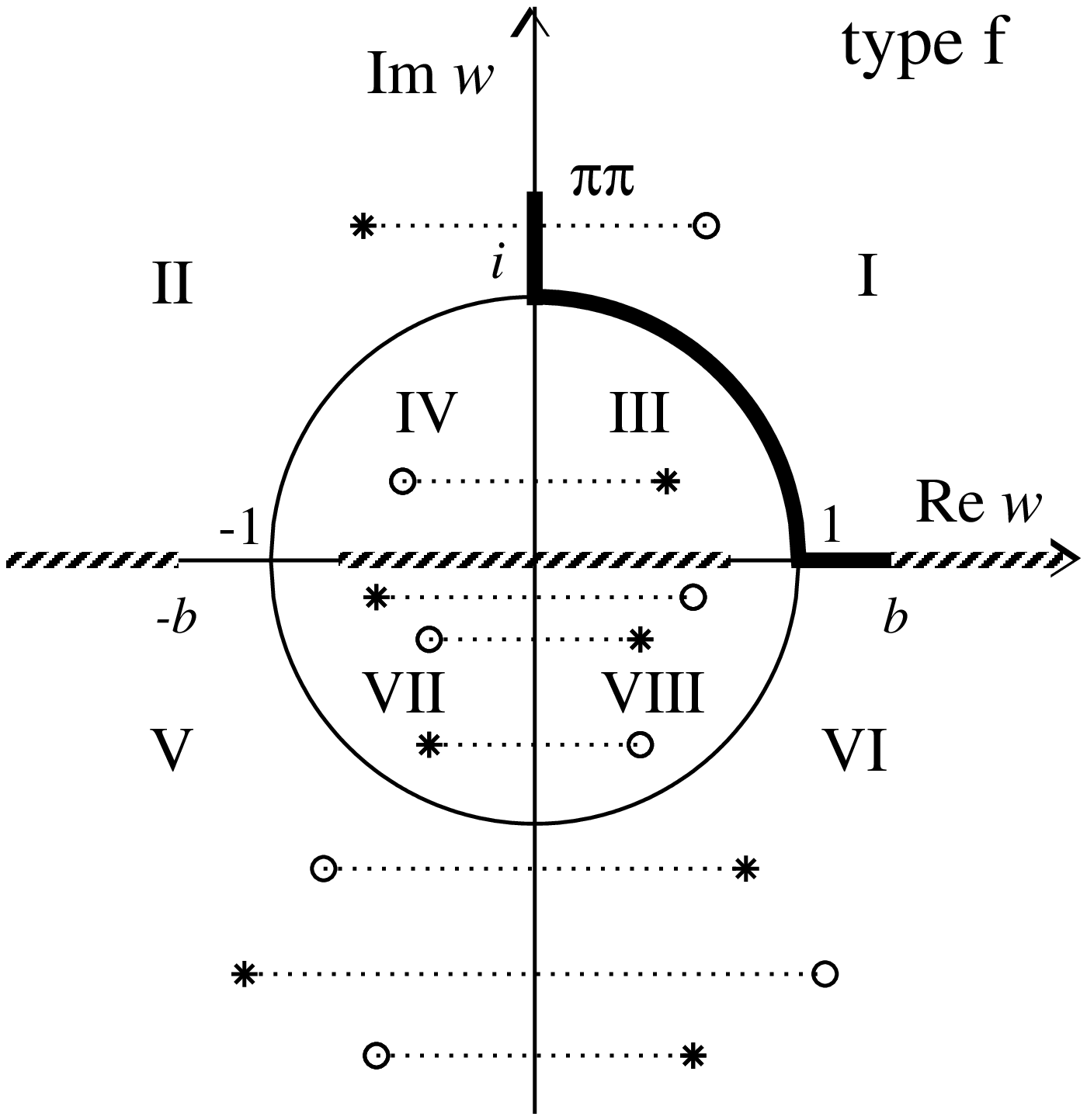}\\
\includegraphics[width=0.4\textwidth,angle=-90]{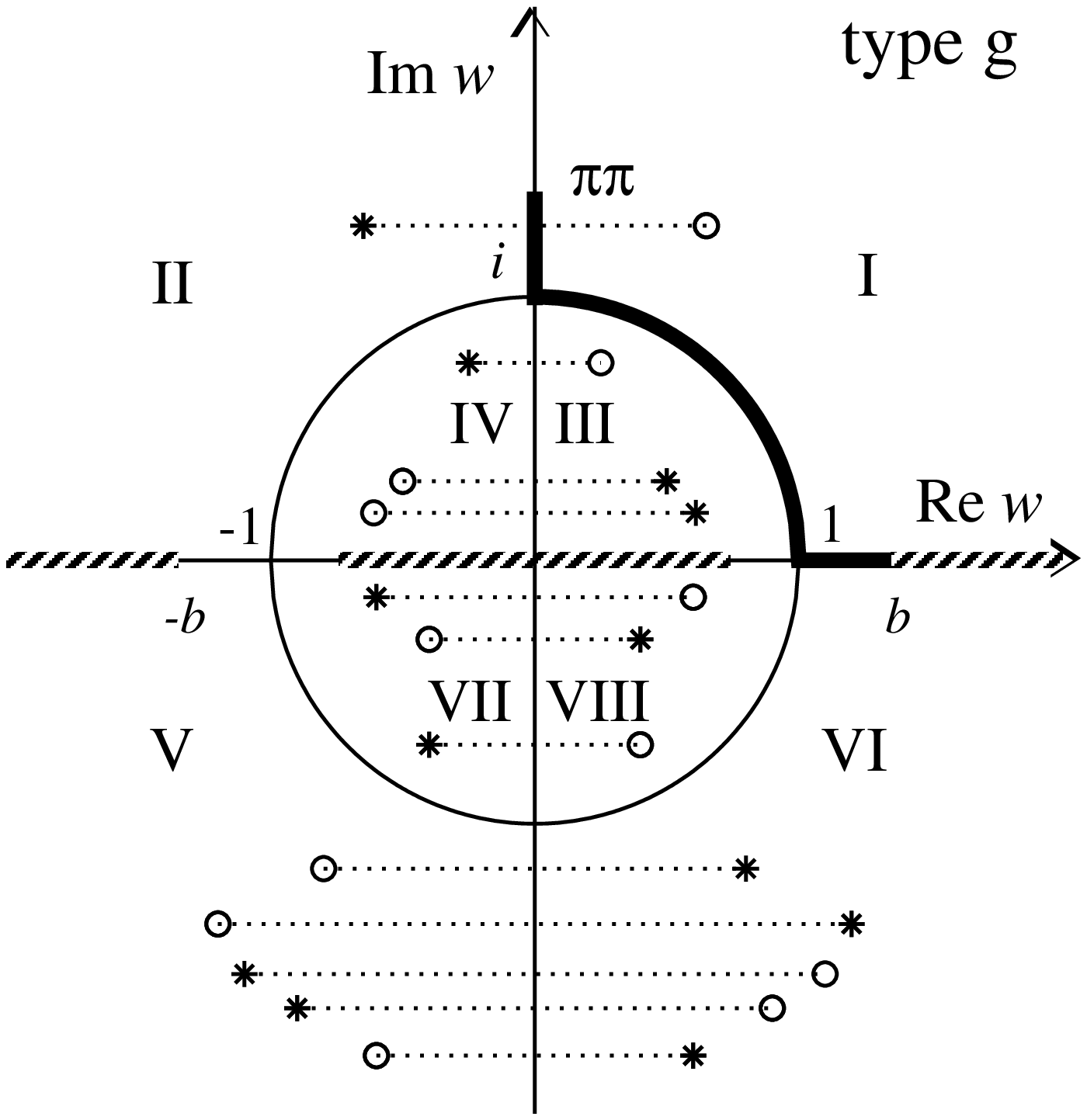}
\end{center}
\vspace*{-0.4cm}
\caption{Representation of resonances of types ({\bf e}), ({\bf f}),
and ({\bf g}). \label{fig:lw_plane_efg}}
\end{figure}

In the corresponding uniformizing variable, we have neglected the $\pi\pi$-threshold branch-point and taken into account the $K\overline{K}$- and $\eta\eta$-threshold branch-points and the left-hand branch-point at $s=0$:
\begin{equation} \label{}
w=\frac{\sqrt{(s-s_2)s_3} + \sqrt{(s-s_3)s_2}}{\sqrt{s(s_3-s_2)}}~~~~(s_2=4m_K^2 ~ {\rm and}~ s_3=4m_\eta^2).
\end{equation}
In Figs.~\ref{fig:lw_plane_abcd} and \ref{fig:lw_plane_efg} we show the representation of resonances of all types ({\bf a}), ({\bf b}),..., ({\bf g}) on the uniformization $w$-plane for the 3-channel-$\pi\pi$-scattering $S$-matrix element.

On the $w$-plane, the Le Couteur--Newton relations are somewhat modified taking account of the used model of initial 8-sheeted Riemann surface (note that on the $w$-plane the points $w_0$, $-w_0^{-1}$, $-w_0$, and $w_0^{-1}$ correspond to the $s$-variable point $s_0$ on sheets I, IV, V, and VIII, respectively):
\begin{equation} \label{}
S_{11}=\frac{d^* (-w^*)}{d(w)},~~~~~~~~
S_{22}=\frac{d(-w^{-1})}{d(w)},~~~~~~~~
S_{33}=\frac{d(w^{-1})}{d(w)},
\end{equation}
\begin{equation} \label{}
S_{11}S_{22}-S_{12}^2=\frac{d^*({w^*}^{-1})}{d(w)},~~~~~~~~
S_{11}S_{33}-S_{13}^2=\frac{d^*(-{w^*}^{-1})}{d(w)},~~
\end{equation}
\begin{equation} \label{}
S_{22}S_{33}-S_{23}^2=\frac{d(-w)}{d(w)}.
\end{equation}
Since the used model Riemann surface means only the consideration of the semi-sheets of the initial Riemann surface nearest to the physical region, then in this case there is no point in saying for the property of the real analyticity of the amplitudes. The 3-channel unitarity requires the following relations to hold for physical $w$-values:
\begin{equation} \label{}
|d(-w^*)|\leq |d(w)|,\quad |d(-w^{-1})|\leq |d(w)|,\quad |d(w^{-1})|\leq
|d(w)|,
\end{equation}
\begin{equation} \label{}
|d({w^*}^{-1})|=|d(-{w^*}^{-1})|=|d(-w)|=|d(w)|.
\end{equation}
The $S$-matrix elements in Le Couteur--Newton relations are taken as the products~~$S=S_B S_{res}$; the main (\underline{model-independent}) contribution of resonances, given by the pole clusters, is included in the resonance part $S_{res}$; possible remaining small (\underline{model-dependent}) contributions of resonances and influence of channels which are not taken explicitly into account in the uniformizing variable are included in the background part $S_B$. The d-function for the resonance part is
\begin{equation} \label{}
d_{res}(w)=w^{-\frac{M}{2}}\prod_{r=1}^{M}(w+w_{r}^*)
\end{equation}
where $M$ is the number of resonance zeros, for the background part is
\begin{equation} \label{}
d_B=\mbox{exp}[-i\sum_{n=1}^{3}\frac{\sqrt{s-s_n}}{2m_n}(\alpha_n+i\beta_n)],
\end{equation}
\begin{equation} \label{}
\alpha_n=a_{n1}+a_{n\sigma}\frac{s-s_\sigma}{s_\sigma}\theta(s-s_\sigma)+
a_{nv}\frac{s-s_v}{s_v}\theta(s-s_v),\nonumber
\end{equation}
$$\beta_n=b_{n1}+b_{n\sigma}\frac{s-s_\sigma}{s_\sigma}\theta(s-s_\sigma)+
b_{nv}\frac{s-s_v}{s_v}\theta(s-s_v)$$ where $s_\sigma$ is the
$\sigma\sigma$ threshold; $s_v$ is the combined threshold of the
$\eta\eta^{\prime}$, $\rho\rho$ and $\omega\omega$ channels.

Formalism for calculating di-meson mass distributions of the decays
$J/\psi\to\phi(\pi\pi, K\overline{K})$ and $V^{\prime}\to V\pi\pi$
(e.g., $\psi(2S)\to J/\psi(\pi\pi)$ and $\Upsilon(2S)\to\Upsilon(1S)\pi\pi$) can be found in Refs. \cite{Morgan_Penn_93,Zou_Bugg_94}. There is assumed that pairs of pseudo-scalar mesons of final states have $I=J=0$ and only they undergo strong interactions, whereas a final vector meson ($\phi$, $V$) acts as a spectator. The amplitudes for decays are related with the scattering amplitudes $T_{ij}$ $(i,j=1-\pi\pi,2-K\overline{K})$ as follows:
\begin{equation} \label{}
F(J/\psi\to\phi\pi\pi)=\sqrt{2/3}~[c_1(s)T_{11}+c_2(s)T_{21}],
\end{equation}
\begin{equation} \label{}
F(J/\psi\to\phi K\overline{K})=\sqrt{1/2}~[c_1(s)T_{12}+c_2(s)T_{22}],
\end{equation}
\begin{equation} \label{}
F(V^{\prime}\to V\pi\pi~(V=\psi,\Upsilon))=[(d_1,e_1)T_{11}+(d_2,e_2)T_{21}]
\end{equation}
where  ~~$c_1=\gamma_{10}+\gamma_{11}s$, ~~$c_2=\alpha_2/(s-\beta_2)+\gamma_{20}+\gamma_{21}s$, ~~and $(d_i,e_i)=(\delta_{i0},\rho_{i0})+(\delta_{i1},\rho_{i1})s$~~ are functions of couplings of the $J/\psi$, $\psi(2S)$ and $\Upsilon(2S)$ to channel~$i$; ~$\alpha_2$, $\beta_2$, $\gamma_{i0}$, $\gamma_{i1}$, $\delta_{i0}$, $\rho_{i0}$, $\delta_{i1}$ and $\rho_{i1}$ are free parameters. The pole term in $c_2$ is an approximation of possible $\phi K$ states, not forbidden by OZI rules when considering quark diagrams of these processes. Obviously this pole should be situated on the real $s$-axis below the $\pi\pi$ threshold. This is an effective inclusion of the effect of so called ``crossed channel final state interactions'' in $J/\psi\to\phi K\overline{K}$, which was studied largely, e.g., in Ref.~\cite{Guo}.

The expressions
$$N|F|^{2}\sqrt{(s-s_i)\bigl(m_\psi^{2}-(\sqrt{s}-m_\phi)^{2}\bigr)\bigl(m_\psi^2-
(\sqrt{s}+m_\phi)^2\bigr)}$$
for $J/\psi\to\phi\pi\pi,\phi K\overline{K}$ (and the analogues ones for $V^{\prime}\to V\pi\pi$) give the di-meson mass distributions. N (a normalization constant to data of the experiments) is 0.7512 for Mark~III, 0.3705 for DM2, 5.699 for BES~II, 1.015 for Mark~II, 0.98 for Crystal Ball(80), 4.3439 for Argus, 2.1776 for CLEO, 1.2011 for CUSB, and 0.0788 for Crystal Ball(85).

\section{The combined 3-channel analysis of data on isoscalar S-wave processes $\pi\pi\to\pi\pi,K\overline{K},\eta\eta$ and on $J/\psi\to\phi(\pi\pi, K\overline{K})$, $\psi(2S)\to J/\psi(\pi\pi)$ and $\Upsilon(2S)\to\Upsilon(1S)\pi\pi$}

For the data on multi-channel $\pi\pi$ scattering we used the results of phase analyses
which are given for phase shifts of the amplitudes $\delta_{\alpha\beta}$ and for the
modules of the $S$-matrix elements
$\eta_{\alpha\beta}=|S_{\alpha\beta}|$ ($\alpha,\beta=1,~2,~3$):
\begin{equation} \label{}
S_{\alpha\alpha}=\eta_{\alpha\alpha}e^{2i\delta_{\alpha\alpha}},~~~~~
S_{\alpha\beta}=i\eta_{\alpha\beta}e^{i\phi_{\alpha\beta}}.
\end{equation}
If below the third threshold there is the 2-channel unitarity then
the relations
\begin{equation} \label{}
\eta_{11}=\eta_{22}, ~~ \eta_{12}=(1-{\eta_{11}}^2)^{1/2},~~
\phi_{12}=\delta_{11}+\delta_{22}
\end{equation}
are fulfilled in this energy region.

For the $\pi\pi$ scattering, the data from the threshold to 1.89~GeV are taken from
Refs.~\cite{expd_pipi}. For $\pi\pi\to K\overline{K}$, practically all the accessible
data are used \cite{expd_pipiKK}. For $\pi\pi\to\eta\eta$, we used data for $|S_{13}|^2$
from the threshold to 1.72~GeV \cite{expd_eta_eta}. For decays
$J/\psi\to\phi\pi\pi,\phi K\overline{K}$ we have taken data from Mark~III , from DM2
and from BES~II \cite{Mark_III}; for $\psi(2S)\to J/\psi(\pi^+\pi^-)$ from Mark~II and
for $\psi(2S)\to J/\psi(\pi^0\pi^0)$ from Crystal Ball Collaborations(80) \cite{Mark_II};
for $\Upsilon(2S)\to\Upsilon(1S)(\pi^+\pi^-,\pi^0\pi^0)$ from Argus, CLEO, CUSB, and
Crystal Ball collaborations(85) \cite{Argus}.

In this combined analyses of the coupled scattering processes and decays, it appears that to achieve a consistency with the PDG tables (to have a narrow state) it is necessary to consider two states in the 1500-MeV region -- the narrow $f_0(1500)$ and
wide $f_0^\prime(1500)$ (which is needed for description of the multi-channel $\pi\pi$ scattering).

We have obtained the following preferable scenarios: the
$f_0(500)$ is described by the cluster of type ({\bf a}); the $f_0(1370)$ and $f_0(1500)$, type ({\bf c}) and $f_0^\prime(1500)$, type ({\bf g}); the $f_0(980)$ is represented only by the pole on sheet~II and shifted pole on sheet~III. However, the $f_0(1710)$ can be described by clusters either of type ({\bf b}) or ({\bf c}). For definiteness, we have taken type~({\bf c}). Parameters of resonances and background are changed very insignificantly in comparison with our analysis \cite{SBKLN-1207_6937} performed without consideration of decays $\psi(2S)\to J/\psi(\pi\pi)$ and $\Upsilon(2S)\to\Upsilon(1S)\pi\pi$ -- confirming our previous results.

Parameters of the coupling functions of the decay particles ($J/\psi$, $\psi(2S)$ and
$\Upsilon(2S)$) to channel~$i$, obtained in the analysis, are~
${\alpha_2,\beta_2} =$ 0.0843, 0.0385,
${\gamma_{10},\gamma_{11},\gamma_{20},\gamma_{21}}=1.1826$, 1.2798, -1.9393, -0.9808,
${\delta_{10},\delta_{11},\delta_{20},\delta_{21}} =$ -0.127, 16.621, 5.983, -57.653,
${\rho_{10},\rho_{11},\rho_{20},\rho_{21}} = $0.405, 47.0963, 1.3352, -21.4343.

There is retained the fact that {\it the di-pion mass distribution of the $J/\psi\to\phi\pi\pi$ decay of the BESIII data from the threshold to about 850~MeV prefers surely the solution with the wider $f_0(500)$} -- B-solution. Therefore further we will discuss mainly the B solution.

Satisfactory combined description of all analyzed processes is obtained with the total
$\chi^2/\mbox{NDF}=568.57/(481-65)\approx1.37$; for the
$\pi\pi$ scattering, $\chi^2/\mbox{NDF}\approx1.15$; for $\pi\pi\to K\overline{K}$, $\chi^2/\mbox{NDF}\approx1.65$; for $\pi\pi\to\eta\eta$, $\chi^2/\mbox{ndp}\approx0.87$; for decays $J/\psi\to\phi(\pi\pi, K\overline{K})$, $\chi^2/\mbox{ndp}\approx1.21$; for $\psi(2S)\to J/\psi(\pi\pi)$, $\chi^2/\mbox{ndp}\approx2.43$; for $\Upsilon(2S)\to\Upsilon(1S)\pi\pi$, $\chi^2/\mbox{ndp}\approx1.01$.

\begin{figure}[!thb]
\begin{center}
\includegraphics[width=0.45\textwidth,angle=0]{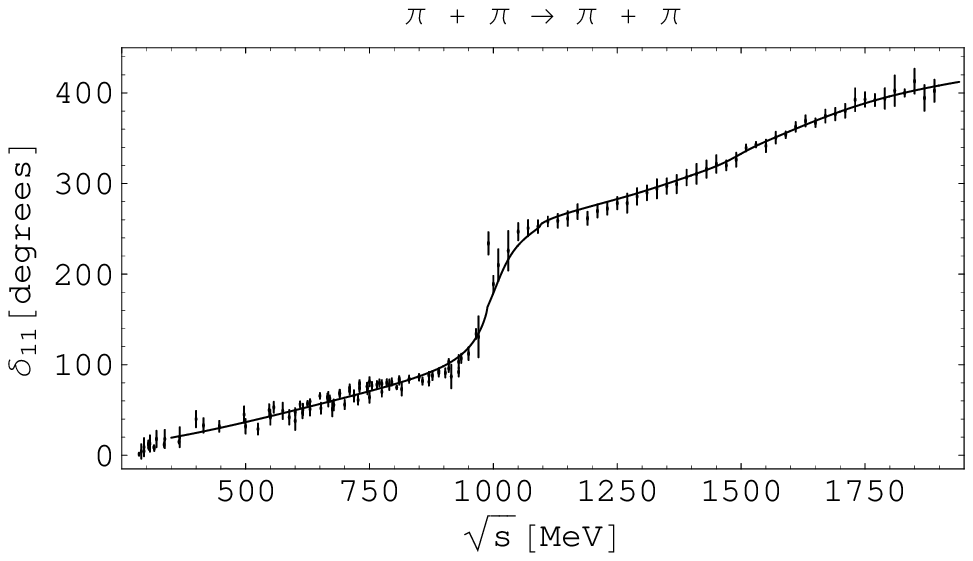}
\includegraphics[width=0.45\textwidth,angle=0]{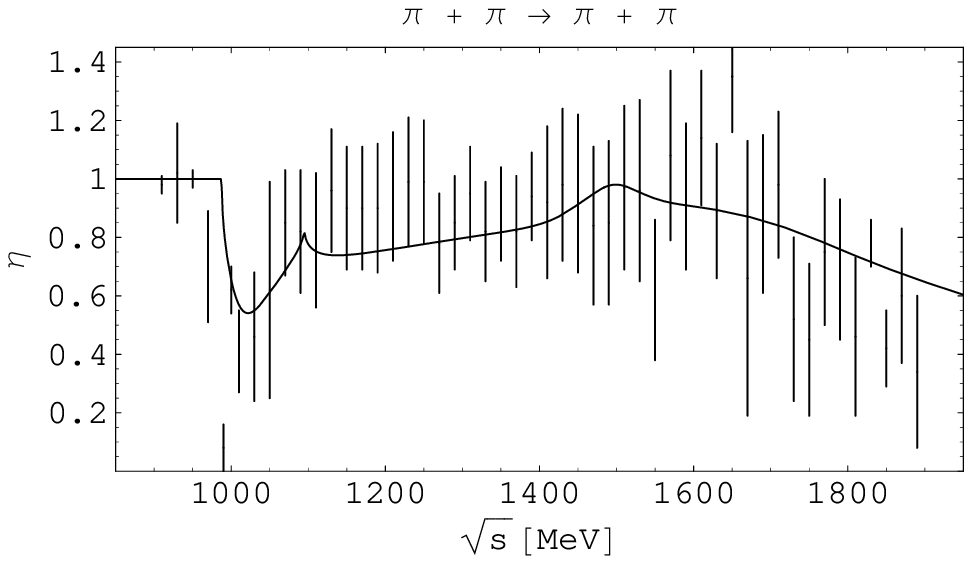}\\
\includegraphics[width=0.45\textwidth,angle=0]{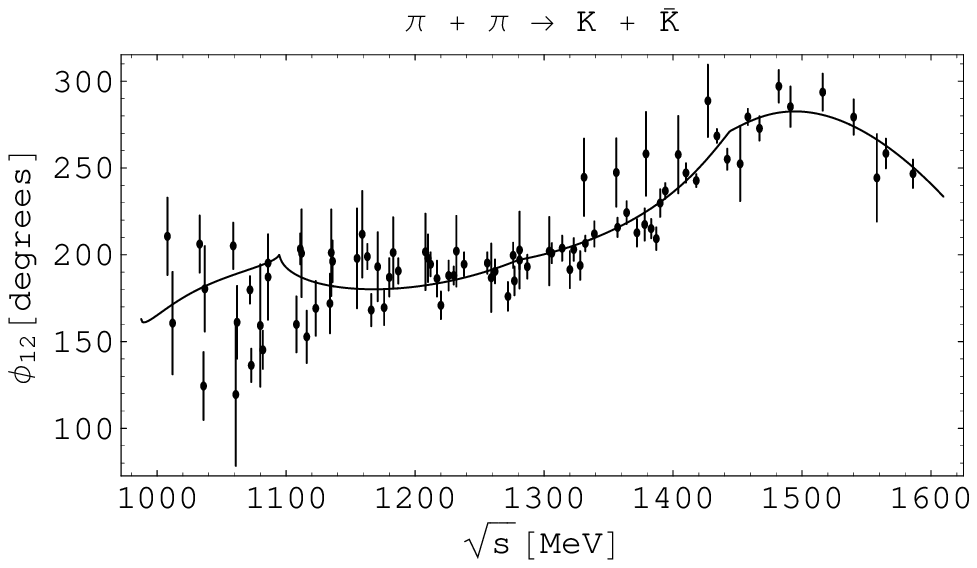}
\includegraphics[width=0.45\textwidth,angle=0]{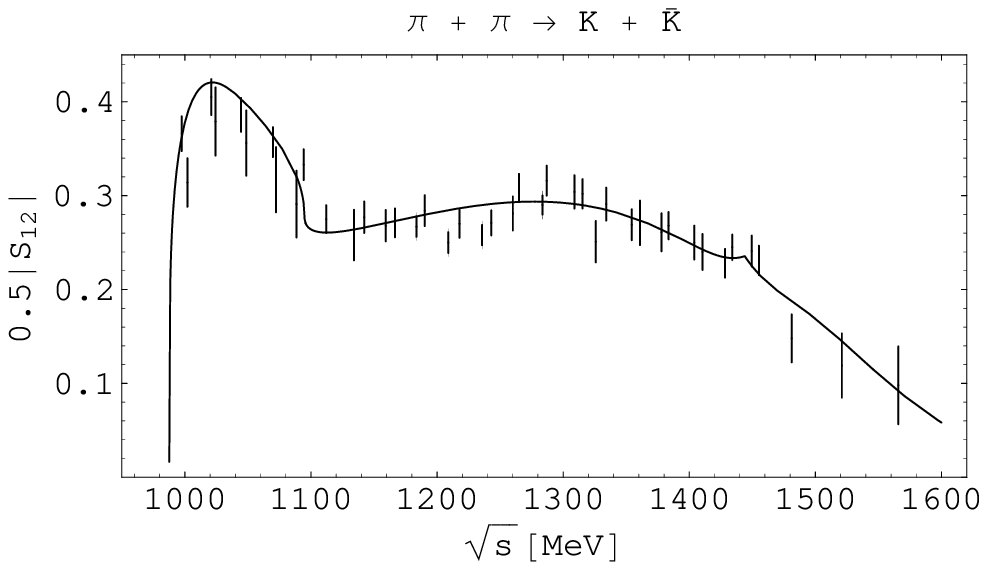}\\
\vspace*{-0.1cm}
\includegraphics[width=0.45\textwidth,angle=0]{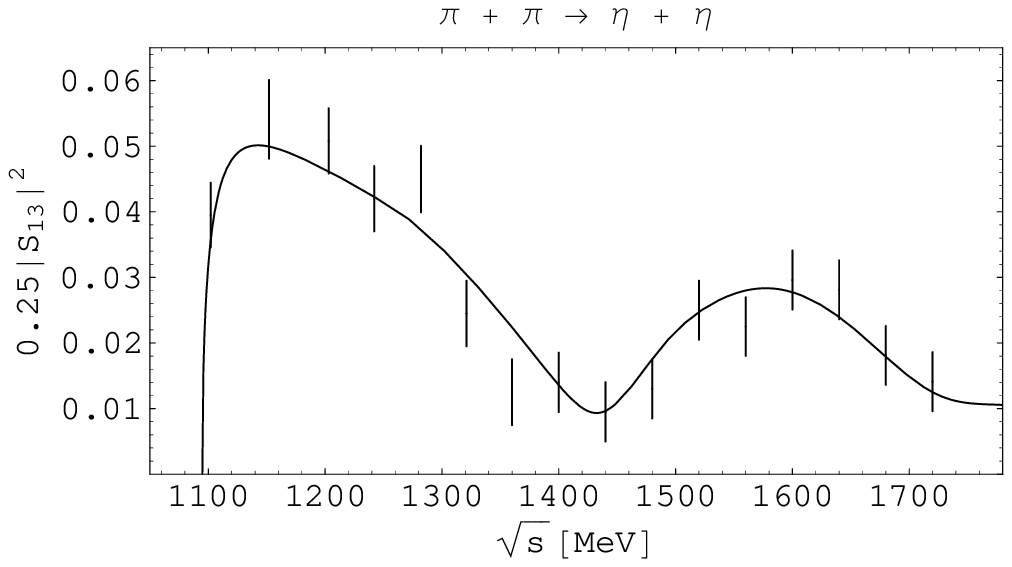}
\vskip -.2cm
\caption{The phase shifts and modules of the $S$-matrix element in the $S$-wave $\pi\pi$-scattering (upper panel, data from \cite{expd_pipi}), in $\pi\pi\to K\overline{K}$ (middle panel, data from \cite{expd_pipiKK}), and the squared module of the $\pi\pi\to\eta\eta$ $S$-matrix element (lower figure, data from \cite{expd_eta_eta}). \label{Fig:multi_pipi}}
\end{center}
\end{figure}

\begin{figure}[!thb]
\begin{center}
\includegraphics[width=0.44\textwidth,angle=0]{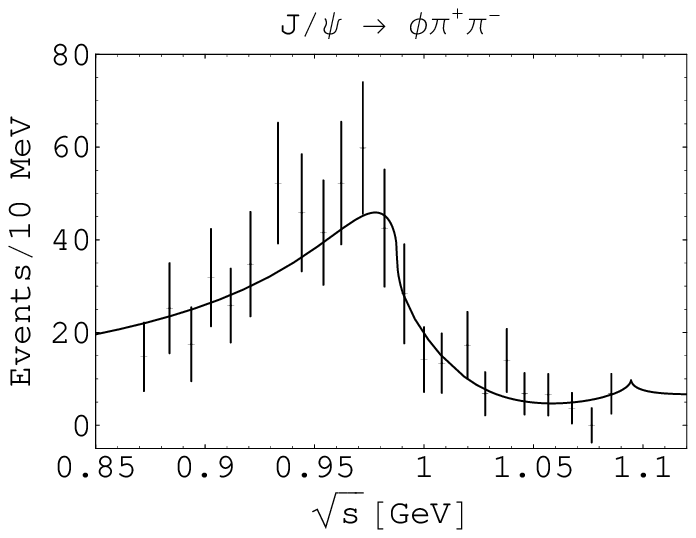}
\includegraphics[width=0.44\textwidth,angle=0]{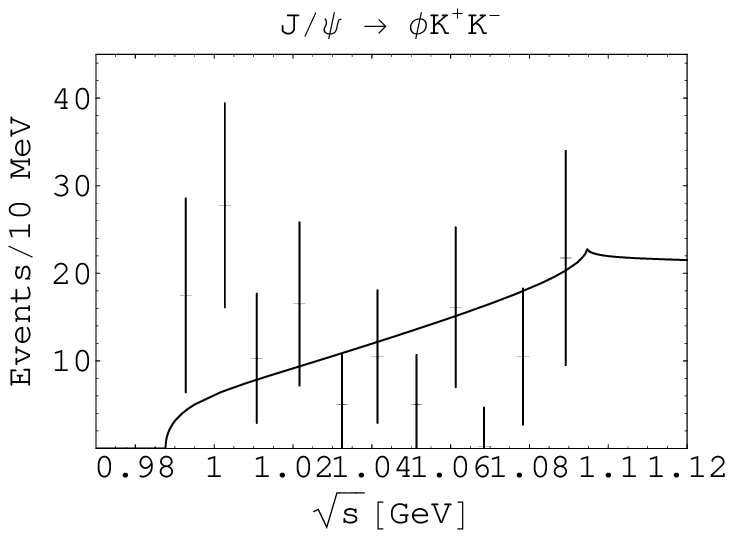}\\
\vspace*{0.1cm}
\includegraphics[width=0.44\textwidth,angle=0]{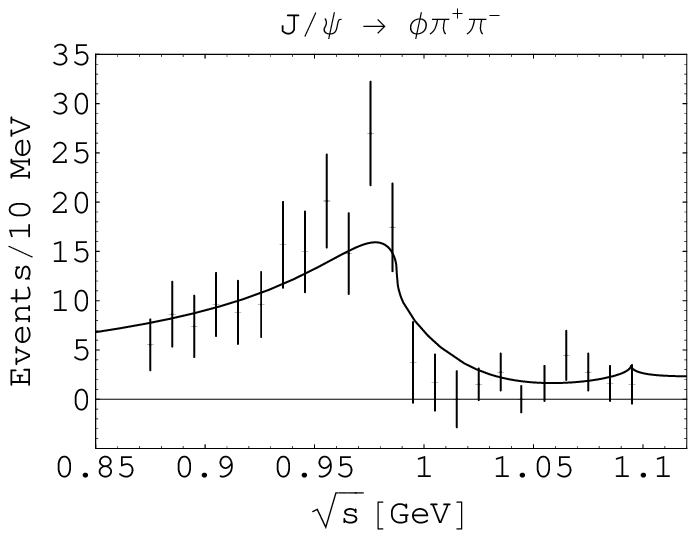}
\includegraphics[width=0.44\textwidth,angle=0]{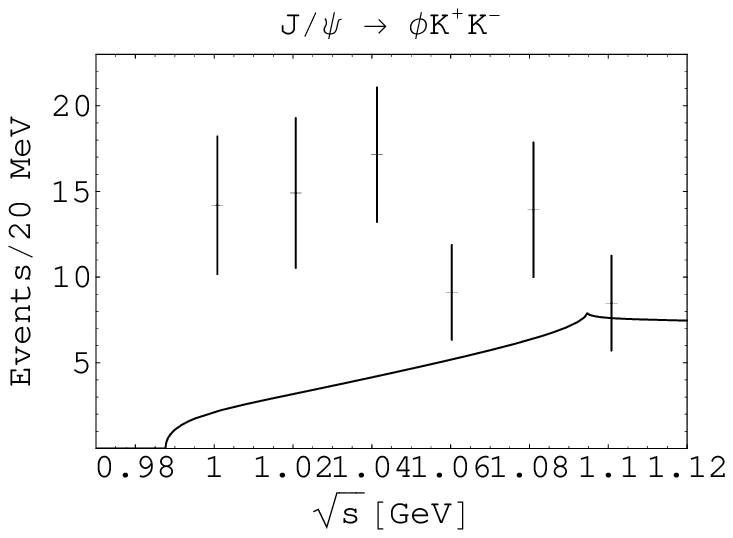}
\vspace*{-0.2cm}\caption{The $J/\psi\to\phi\pi\pi,\phi
K\overline{K}$ decays. The upper panel shows the fit to data of
Mark~III, the lower to DM2. \label{Fig:psi_MarkIII_DM2}}
\end{center}
\end{figure}

\begin{figure}[!thb]
\begin{center}
\includegraphics[width=0.70\textwidth,angle=0]{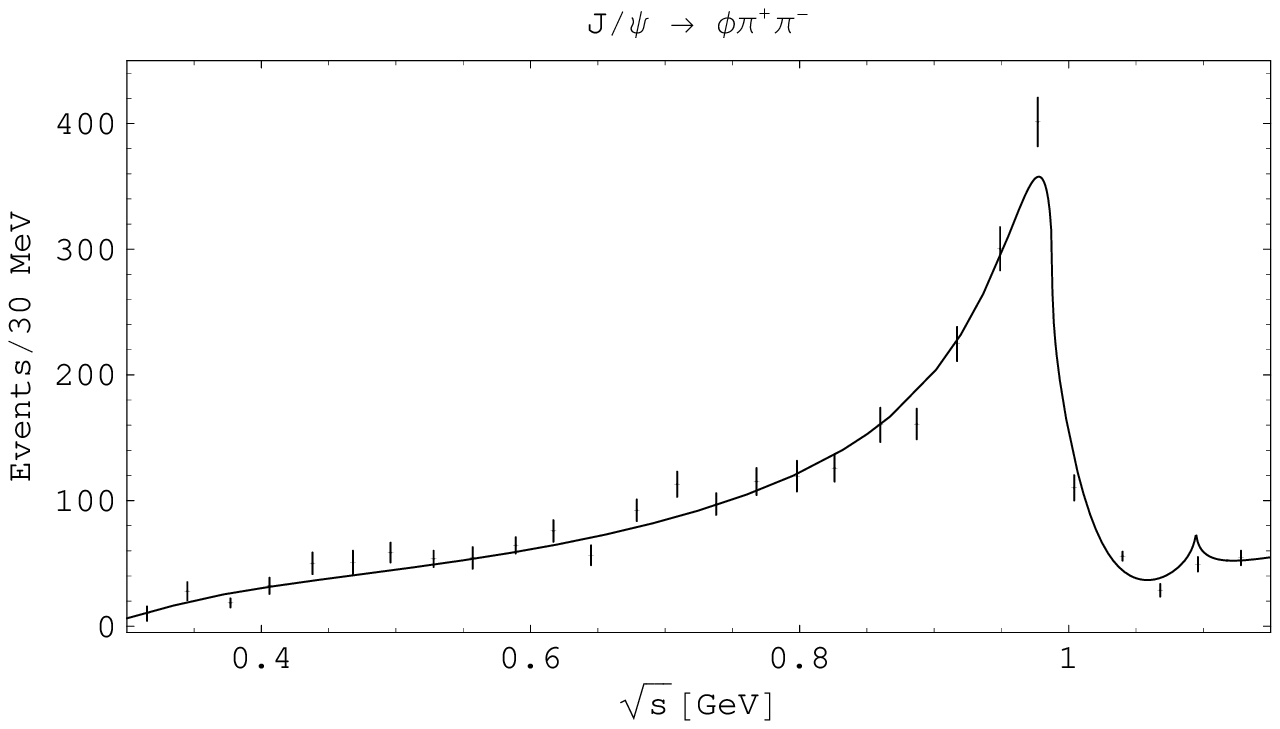}
\vspace*{-0.2cm}\caption{The $J/\psi\to\phi\pi\pi$ decay;
the data of BES~II collaboration. \label{Fig:psi_BESIII}}\end{center}
\end{figure}

\begin{figure}[!thb]
\begin{center}
\includegraphics[width=0.44\textwidth,angle=0]{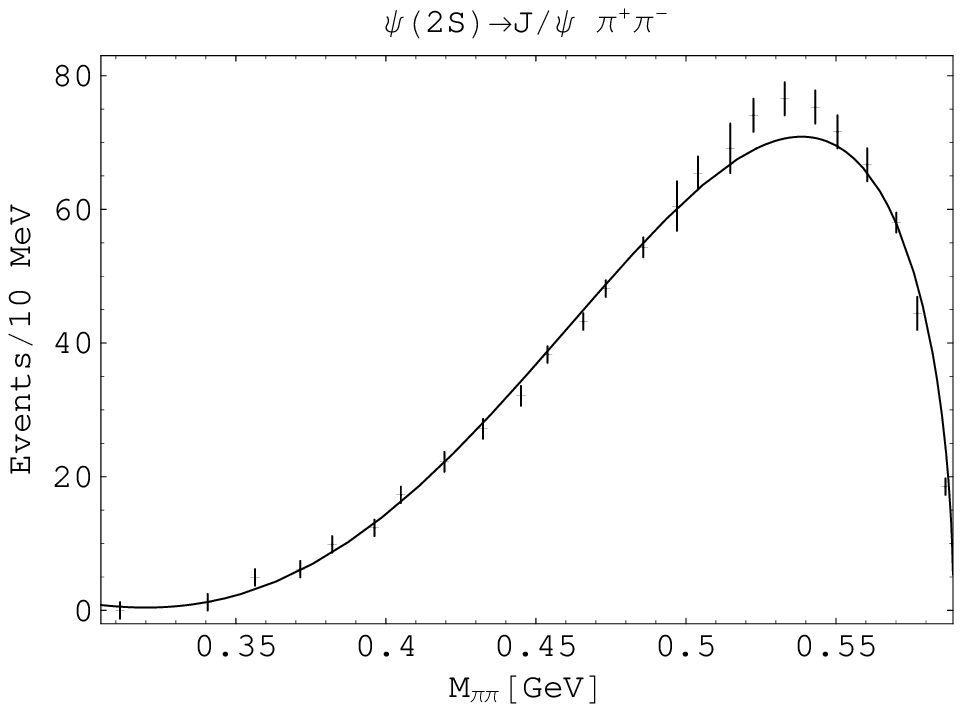}
\includegraphics[width=0.44\textwidth,angle=0]{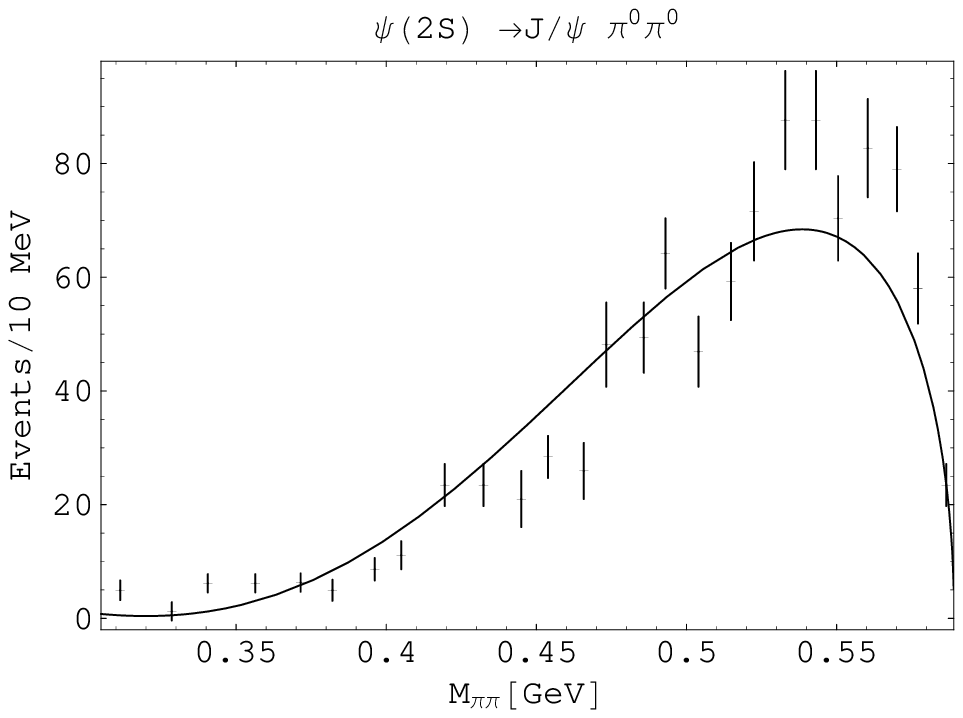}\\
\vspace*{-0.2cm}\caption{The $\psi(2S)\to J/\psi\pi\pi$ decays. The left figure shows the fit to the Mark~II data, the right to Crystal Ball(80). \label{Fig:psi(2S)_decays}}
\end{center}
\end{figure}

\begin{figure}[!thb]
\begin{center}
\includegraphics[width=0.44\textwidth,angle=0]{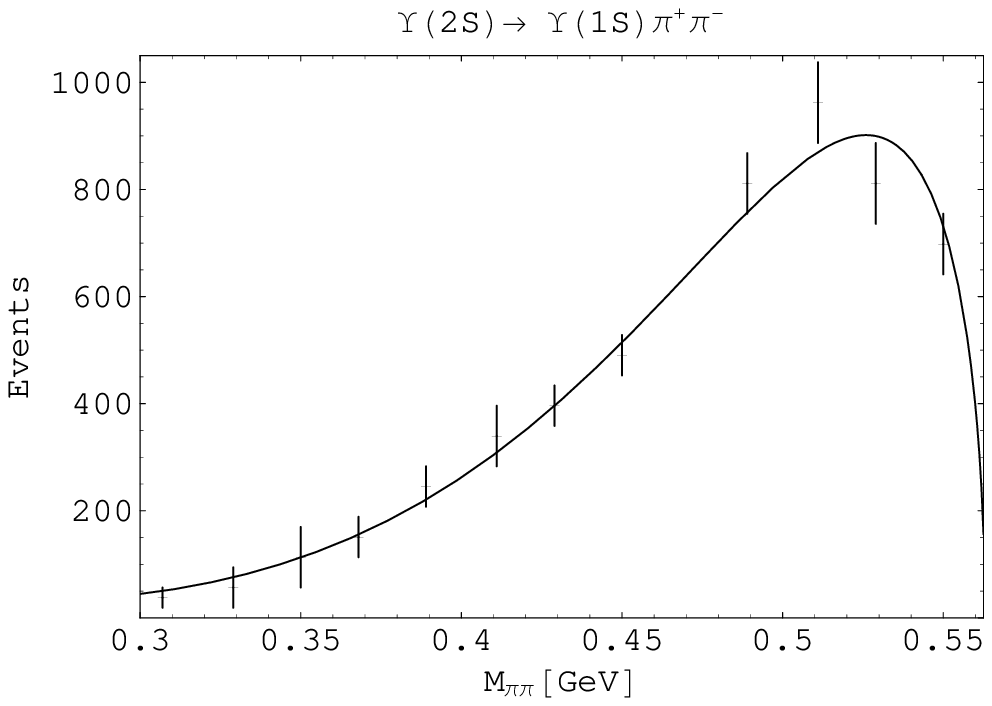}
\includegraphics[width=0.44\textwidth,angle=0]{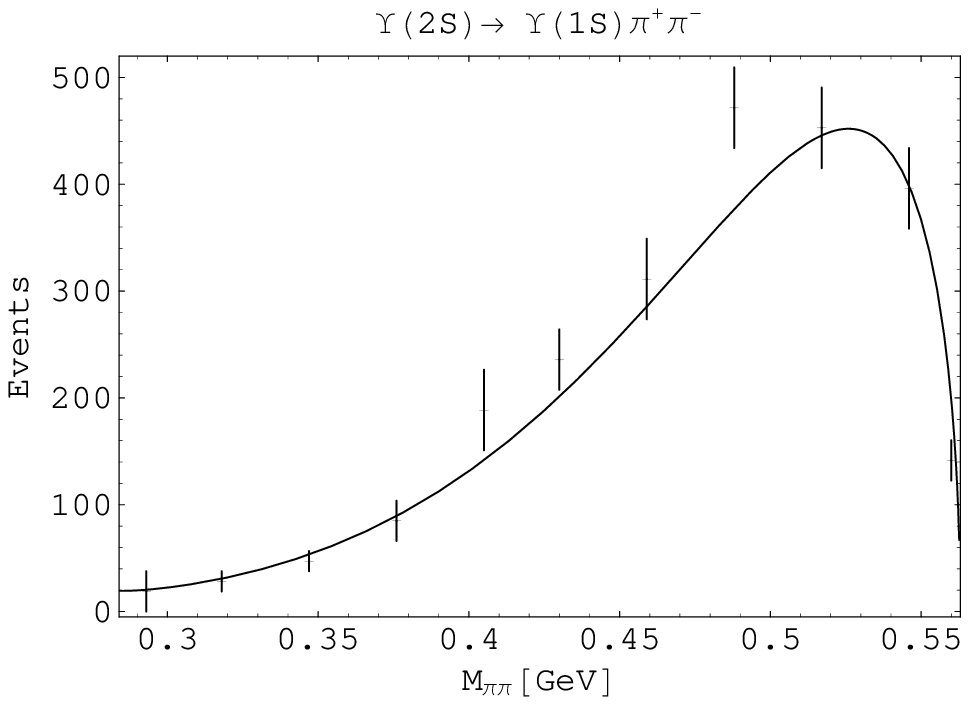}\\
\vspace*{0.1cm}
\includegraphics[width=0.44\textwidth,angle=0]{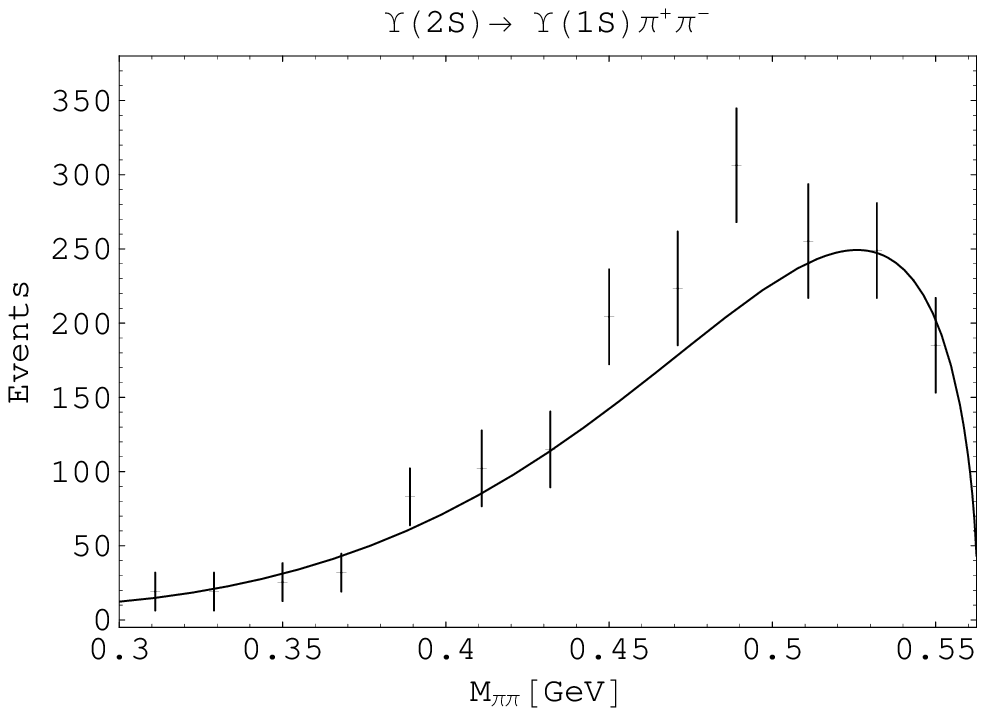}
\includegraphics[width=0.44\textwidth,angle=0]{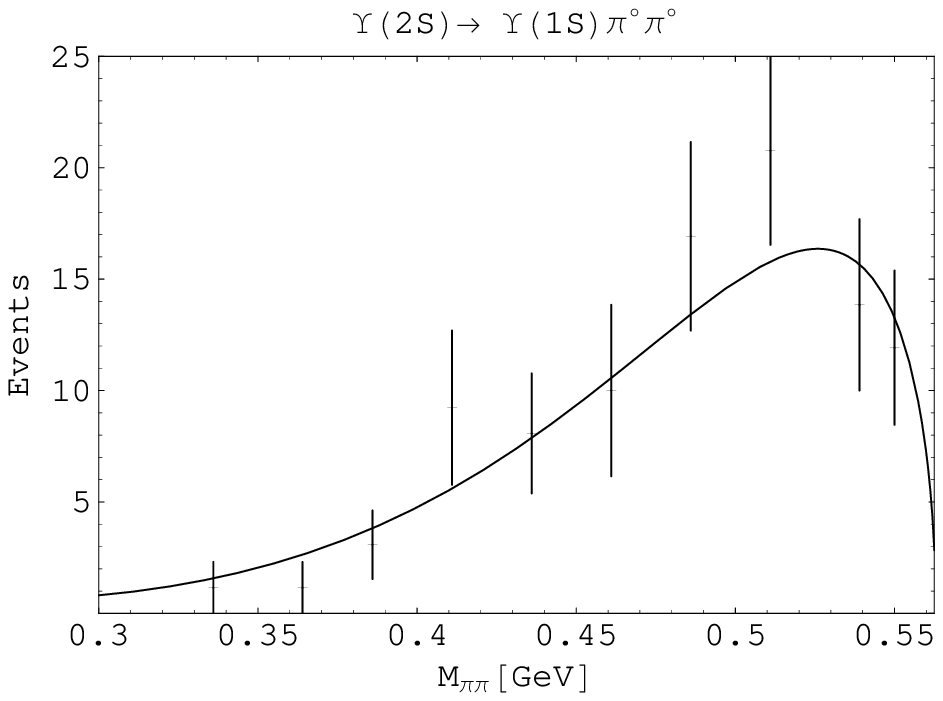}
\vspace*{-0.2cm}
\caption{The $\Upsilon(2S)\to\Upsilon(1S)\pi\pi$ decays.
The upper panel shows the fit to data of Argus (left) and CLEO (right),
the lower to CUSB (left) and Crystal Ball (85) (right). \label{Fig:Ups(2S)_decays}}\end{center}
\end{figure}

In Figs.~\ref{Fig:multi_pipi}--\ref{Fig:Ups(2S)_decays}, we show results of fitting to the experimental data.

The di-pion mass distribution in decay $J/\psi\to\phi\pi\pi$, obtained the BES~II
collaboration and having rather small errors (Fig.~\ref{Fig:psi_BESIII}), rejects
dramatically the A solution with the narrower $f_0(600)$. The corresponding curve
lies considerably below the data from the threshold to about 850~MeV .

The obtained background parameters are:\\
\underline{$a_{11}=0.0$, $a_{1\sigma}=0.0199$, $a_{1v}=0.0$,} \underline{$b_{11}=b_{1\sigma}=0.0$,
$b_{1v}=0.0338$,} $a_{21}=-2.4649$, $a_{2\sigma}=-2.3222$, $a_{2v}=-6.611$,
$b_{21}=b_{2\sigma}=0.0$, $b_{2v}=7.073$, $b_{31}=0.6421$, $b_{3\sigma}=0.4851$,
$b_{3v}=0$; $s_\sigma=1.6338~{\rm GeV}^2$, $s_v=2.0857~{\rm GeV}^2$.

The very simple description of the $\pi\pi$-scattering background (underlined values) confirms well our assumption \underline{$S=S_B S_{res}$} and also that representation of multi-channel resonances by the pole clusters on the uniformization plane is good and quite sufficient. Moreover, this shows that {\it the consideration of the left-hand branch-point at $s=0$ in the uniformizing variable solves partly a problem of some approaches (see, e.g., \cite{Achasov-Shest}) that the wide-resonance parameters are strongly controlled by the non-resonant background.} Note also that the zero background of the $\pi\pi$ scattering, in addition to the fact that $f_0(500)$ is described by the cluster, indicates this state to be the resonance (not a dynamically generated state). The point is that after the account of the left-hand branch-point at $s=0$ remaining contributions of the crossed $u$- and $t$-channels are meson exchanges.
The elastic background of the $\pi\pi$ scattering is related mainly to
contributions of the crossed channels. Its zero value means that the exchange by nearest $\rho$-meson is obliterated by the exchange by a
particle of near mass contributing with opposite sign (the scalar $f_0(500)$) \cite{SKN-epja}.

In Table~\ref{tab:clusters} the obtained pole-clusters for the $f_0$ resonances are shown on the $\sqrt{s}$-plane.
\begin{table}[!htb]
{
\def\arraystretch{1.3}
\begin{tabular}{|c|c|c|c|c|c|c|c|}
\hline ${\rm Sheet}$ & {} & $f_0(500)$ & $f_0(980)$ & $f_0(1370)$ & $f_0(1500)$ & $f_0^\prime(1500)$ & $f_0(1710)$ \\ \hline
II & {${\rm E}_r$} & $514.5\pm12.4$ & $1008.1\pm3.1$ & {} & {} & $1512.7\pm4.9$ & {} \\
{} & {$\Gamma_r/2$} & $465.6\pm5.9$ & $32.0\pm1.5$ & {} & {} & $285.8\pm12.9$ & {} \\
\hline III & {${\rm E}_r$} & $544.8\pm17.7$ & $976.2\pm5.8$ & $1387.6\pm24.4$ & {} & $1506.2\pm9.0$ & {} \\
{} & {$\Gamma_r/2$} & $465.6\pm5.9$ & $53.0\pm2.6$ & $166.9\pm41.8$ & {} & $127.9\pm10.6$ & {} \\
\hline IV & {${\rm E}_r$} & {} & {} & 1387.6$\pm$24.4 & {} & 1512.7$\pm$4.9 & {} \\
{} & {$\Gamma_r/2$} & {} & {} & $178.5\pm37.2$ & {} & $216.0\pm17.6$ & {} \\
\hline V & {${\rm E}_r$} & {} & {} & 1387.6$\pm$24.4 & $1493.9\pm3.1$ & $1498.9\pm7.2$ & $1732.8\pm43.2$ \\
{} & {$\Gamma_r/2$} & {} & {} & $260.9\pm73.7$ & $72.8\pm3.9$ & $142.2\pm6.0$ & $114.8\pm61.5$ \\
\hline VI & {${\rm E}_r$} & $566.5\pm29.1$ & {} & 1387.6$\pm$24.4 & $1493.9\pm5.6$
& $1511.4\pm4.3$ & 1732.8$\pm$43.2 \\
{} & {$\Gamma_r/2$} & $465.6\pm5.9$ & {} & $249.3\pm83.1$ & $58.4\pm2.8$ & $179.1\pm4.0$ & $111.2\pm8.8$ \\
\hline VII & {${\rm E}_r$} & $536.2\pm25.5$ & {} & {} & $1493.9\pm5.0$ & $1500.5\pm9.3$ & 1732.8$\pm$43.2 \\
{} & {$\Gamma_r/2$} & $465.6\pm5.9$ & {} & {} & $47.8\pm9.3$ & $99.7\pm18.0$ & $55.2\pm38.0$ \\
\hline VIII & {${\rm E}_r$} & {} & {} & {} & $1493.9\pm3.2$ & 1512.7$\pm$4.9 & 1732.8$\pm$43.2 \\
{} & {$\Gamma_r/2$} & {} & {} & {} & $62.2\pm9.2$ & $299.6\pm14.5$ & $58.8\pm16.4$ \\
\hline
\end{tabular}}
\caption{The pole clusters for resonances on the $\sqrt{s}$-plane.
~$\sqrt{s_r}\!=\!{\rm E}_r\!-\!i\Gamma_r/2$~[MeV]. \label{tab:clusters}}
\end{table}
Generally, {\it wide multi-channel states are most adequately represented by pole clusters}, because the pole clusters give the main model-independent effect of resonances. The pole positions are rather stable characteristics for various models, whereas masses and widths are very model-dependent for wide resonances.
However, mass values are needed in some cases, e.g., in mass relations for multiplets. Therefore, we stress that such parameters of the wide multi-channel states, as {\it masses, total widths and coupling constants with channels, should be calculated using the poles on sheets II, IV and VIII}, because only on these sheets the analytic continuations have the forms: $$\propto 1/S_{11}^{\rm I},~~
\propto 1/S_{22}^{\rm I}~~{\rm and}~~\propto 1/S_{33}^{\rm I},$$
respectively, i.e., the pole positions of resonances are at the same points of the complex-energy plane, as the resonance zeros on the physical sheet, and are not shifted due to the coupling of channels.

It appears that neglecting the above-indicated principle can cause
misunderstandings. This concerns especially the analyses which do not consider the structure of the Riemann surface of the $S$-matrix. For example, in literature there is a common opinion (delusion) that the resonance parameters should be calculated using resonance poles nearest to the physical region. This is right only in the one-channel case. In the multi-channel case this is not correct. It is obvious that, e.g., the resonance pole on sheet~III, which is situated above the second threshold, is nearer to the physical region than the pole on sheet~II
from the pole cluster of the same resonance since above the $K\overline{K}$ threshold the physical region (an upper edge of the right-hand cut) is joined directly with sheet~III. Therefore, the pole on sheet~III influences most strongly on the energy behaviour of the amplitude and this pole will be found in the analyses, not taking into account the structure of the Riemann surface and the representation of resonances by the pole clusters.

E.g., if the resonance part of amplitude is taken as
\begin{equation} \label{}
T^{res}=\sqrt{s}~\Gamma_{el}/(m_{res}^2-s-i\sqrt{s}~\Gamma_{tot}),
\end{equation}
for the mass and total width, one obtains
\begin{equation} \label{}
m_{res}=\sqrt{{\rm E}_r^2+\left(\Gamma_r/2\right)^2}~~~
{\rm and}~~~\Gamma_{tot}=\Gamma_r,
\end{equation}
where the pole position $\sqrt{s_r}\!=\!{\rm E}_r\!-\!i\Gamma_r/2$ must be taken on sheets II, IV, VIII, depending on the resonance classification. In Table~\ref{tab:masses_widths} the obtained values are given.
\begin{table}[htb!]
\def\arraystretch{1.4}
\centerline{\begin{tabular}{|c|c|c|c|c|c|c|}
\hline {} & $f_0(500)$ & $f_0(980)$ & $f_0(1370)$ & $f_0(1500)$ & $f_0^\prime(1500)$ & $f_0(1710)$\\ \hline
$m_{res}$[MeV] & 693.9$\pm$10.0 & 1008.1$\pm$3.1 & 1399.0$\pm$24.7 & 1495.2$\pm$3.2 & 1539.5$\pm$5.4 & 1733.8$\pm$43.2 \\ \hline
$\Gamma_{tot}$[MeV] & 931.2$\pm$11.8 & 64.0$\pm$3.0
& 357.0$\pm$74.4 & 124.4$\pm$18.4 & 571.6$\pm$25.8 & 117.6$\pm$32.8 \\
\hline
\end{tabular}}
\caption{The masses and total widths of the $f_0$ resonances. \label{tab:masses_widths}}
\end{table}

\section{Discussion and conclusions}
\begin{itemize}
\item
 In the combined analysis of data on isoscalar S-wave processes
 $\pi\pi\to\pi\pi,K\overline{K},\eta\eta$ and on decays
 $J/\psi\to\phi(\pi\pi, K\overline{K})$, $\psi(2S)\to J/\psi(\pi\pi)$  and $\Upsilon(2S)\to\Upsilon(1S)\pi\pi$ from the Argus, Crystal Ball, CLEO, CUSB, DM2, Mark~II, Mark~III, and BES~II collaborations, an additional confirmation of the $f_0(500)$ with mass about 700~MeV and width 930~MeV is obtained. This mass value accords with prediction ($m_{\sigma}\approx m_\rho$) on the basis of mended symmetry by Weinberg \cite{Wei90} and with a refined analysis using the large-$N_c$ consistency conditions between the unitarization and resonance saturation suggesting $m_\rho-m_\sigma=O(N_c^{-1})$ \cite{Nieves-Arriola}. Also, e.g., the prediction of a soft-wall AdS/QCD approach~\cite{GLSV_13} for the mass of the lowest $f_0$ meson -- 721~MeV -- practically coincides with the value obtained in our work.

Note that for the $f_0(500)$, the found pole position on sheet~II is $514.5\pm12.4-i(465.6\pm5.9)$ MeV. The real part is in a good agreement with the results of other analyses cited in the PDG tables of 2012: The PDG estimation for the $f_0(500)$ pole is $400\div550-i(200\div350)$ MeV.
The obtained imaginary part is larger than that given in these other analyses. The above discussion concerns solution B.
The imaginary part of $f_0(500)$ pole in solution A (343 MeV) \cite{SBL-PRD12} is in agreement with the PDG estimation. However, solution A is inconsistent to data on the $J/\psi\to\phi\pi\pi$ decay from BES II collaboration: The corresponding curve in Fig.~\ref{Fig:psi_BESIII} lies considerably below the data from the threshold to about 850 MeV. Therefore, solution A is not considered
in this paper. A partial explanation why in other works one obtains smaller width of the $f_0(500)$ (in comparison with our result) was given in Refs.~\cite{SBKLN-1206_3438,SBKLN-PRD12}. Anyway there stays a question of too large width of the $f_0(500)$.
One can suppose that we observe a superposition of two states -- the $\sigma$-meson and a dynamically generated $4\pi$ state.

\item
Indication for $f_0(980)$ is obtained to be a non-$q{\bar q}$ state, e.g., the bound $\eta\eta$ state, because this state lies slightly above the $K\overline{K}$ threshold and is described by the pole on sheet II and by the shifted pole on sheet III without the corresponding (for standard clusters) poles on sheets VI and VII. The obtained parameters for $f_0(980)$ are $m_{res}=1008.1\pm3.1$~MeV and $\Gamma_{tot}=64\pm 3$~MeV. For the popular (some time ago) interpretation of the $f_0(980)$ as a $K\overline{K}$ molecule \cite{Isgur,Jansen,BGL} it was important that the mass value of this state was below the $K\overline{K}$ threshold. In the PDG tables of 2010 its mass is 980$\pm$10~MeV. We found in all combined analyses of the multi-channel $\pi\pi$ scattering the $f_0(980)$  slightly above 1~GeV, as in the dispersion-relations analysis only of the $\pi\pi$ scattering \cite{GarciaMKPRE-11}. In the PDG tables of 2012, for the mass of $f_0(980)$ an important alteration appeared: now there is given the estimation 990$\pm$20~MeV.

\item
The ${f_0}(1370)$  and $f_0 (1710)$ have the dominant $s{\bar s}$ component. The conclusion
about the ${f_0}(1370)$ agrees quite well with that by the Crystal Barrel
Collaboration~\cite{Amsler95} where the ${f_0}(1370)$ is identified as $\eta\eta$ resonance
in the $\pi^0\eta\eta$ final state of the ${\bar p}p$ annihilation. This explains also quite
well why one did not find this state considering only the $\pi\pi$
scattering~\cite{MO-02-2,Ochs10}. The conclusion about the $f_0(1710)$ is consistent with
the experimental facts that this state is observed in
$\gamma\gamma\to K_SK_S$~\cite{Braccini99} but not in
$\gamma\gamma\to\pi^+\pi^-$ \cite{Barate00}.

\item
In the 1500-MeV region, there are two states: the $f_0(1500)$ ($m_{res}\approx1495$~MeV, $\Gamma_{tot}\approx124$~MeV) and the $f_0^\prime(1500)$ ($m_{res}\approx1539$~MeV, $\Gamma_{tot}\approx574$~MeV). The $f_0^\prime(1500)$ is interpreted as a glueball taking into account its biggest width among the enclosing states \cite{Anisovich97}. As to the large width of glueball, it is worth to indicate Ref.~\cite{Ellis-Lanik}. There an effective QCD Lagrangian with the broken scale and chiral symmetry is used, where a glueball is introduced to theory as a dilaton and its existence is related to breaking of scale symmetry in QCD. The $\pi\pi$ decay width of the glueball, estimated using low-energy theorems, is $\Gamma(G\to\pi\pi)\approx 0.6\,{\rm GeV}\times(m_G/1\,{\rm GeV})^5$ where $m_G$ is the glueball mass. I.e., if the glueball with the mass about 1~GeV exists, then its width would be near 600 MeV. Of course, the use of the above formula is doubtful above 1~GeV, however, a trend for the glueball to be wide is apparently seen. On the other hand, a two-flavour linear sigma model with global chiral symmetry and (axial-)vector mesons as well as an additional glueball degree of freedom where the glueball is also introduced as a dilaton \cite{Parganlija-glbl}, there arises the rather narrow resonance in the 1500-MeV region as predominantly a glueball with a subdominant $qq$ component. On second thoughts, this result can be considered as preliminary due to using a quite rough flavor-symmetry SU($N_f = 3$) in the calculations or, e.g., evaluating the $4\pi$ decay, the intermediate state consisting of two $f_0(500)$ mesons is not included. In Ref.~\cite{GGLF}, where the two-pseudoscalar and two-photon decays of the scalars between 1--2~GeV were analyzed in the framework of a chiral Lagrangian and the glueball was included as a flavor-blind composite mesonic field, the glueball
was found to be rather narrow.

\item
Taking into account the discovery of isodoublet $K_0^*(800)$ \cite{PDG12} (see also \cite{SBGL-PPN10}), we propose the following assignment of the scalar mesons to lower nonets, excluding
the $f_0(980)$ as the non-$q{\bar q}$ state. The lowest nonet: the isovector $a_0(980)$, the isodoublet $K_0^*(900)$, and $f_0(500)$ and $f_0(1370)$ as mixtures of the 8th component of octet and the SU(3) singlet. The Gell-Mann--Okubo (GM-O) formula
$$3m_{f_8}^2=4m_{K_0^*}^2-m_{a_0}^2$$ gives~ $m_{f_8}=870$~MeV. In relation for masses of
nonet $$m_\sigma+m_{f_0(1370)}=2m_{K_0^*(900)}$$ the left-hand side is by about 14\% bigger than the right-hand one.

\item
For the next nonet we find: the isovector $a_0(1450)$, the isodoublet
$K_0^*(1450)$, and two isoscalars $f_0(1500)$ and $f_0(1710)$. From the GM-O formula, $m_{f_8}\approx1450$~MeV. In formula $$m_{f_0(1500)}+m_{f_0(1710)}=2m_{K_0^*(1450)}$$ the left-hand side is by about 10\% bigger than the right-hand one.

\item
This assignment removes a number of questions, stood earlier when placing the scalar mesons to nonets, and does not put any new. The mass formulas indicate to non-simple mixing scheme. The breaking of the last two mass relations tells us that the $\sigma\!-\!f_0(1370)$ and $f_0(1500)\!-\!f_0(1710)$ systems get additional contributions absent in the $K_0^*(900)$ and $K_0^*(1450)$, respectively. A search of the adequate mixing scheme is complicated by the fact that here there is also a remaining chiral symmetry, though, on the other hand, this permits one to predict correctly, {\it e.g.}, the $\sigma$-meson mass \cite{Wei90}.

\end{itemize}

\section*{Acknowledgments}

This work was supported in part by the Grant Program of Plenipotentiary
of Slovak Republic at JINR, the Heisenberg-Landau Program, the
Votruba-Blokhintsev Program for Cooperation of Czech Republic with JINR,
the Grant Agency of the Czech Republic (grant No. P203/12/2126),
the Bogoliubov-Infeld Program for Cooperation of Poland with JINR,
the DFG under Contract No. LY 114/2-1.
The work was also partially supported under the project 2.3684.2011 of
Tomsk State University.


\begin{thebibliography}{99}

\bibitem{PDG12}
J.~Beringer et al. (PDG), Phys. Rev. {\bf D86} 010001 (2012).

\bibitem{SBKN-PRD10}
Yu.S.~Surovtsev, P.~Byd\v{z}ovsk\'y, R.~Kami\'nski, and M.~Nagy,
Phys. Rev. {\bf D81} 016001 (2010).

\bibitem{SBL-PRD12}
Yu.S.~Surovtsev, P.~Byd\v{z}ovsk\'y and V.E.~Lyubovitskij,
Phys. Rev. {\bf D85} 036002 (2012).

\bibitem{SBKLN-PRD12}
Yu.S.~Surovtsev, P.~Byd\v{z}ovsk\'y, R.~Kami\'nski, V.E.~Lyubovitskij, M.~Nagy,
Phys. Rev. {\bf D86} 116002 (2012).

\bibitem{SBKLN-1206_3438}
Yu.S.~Surovtsev, P.~Byd\v{z}ovsk\'y, R.~Kami\'nski, V.E.~Lyubovitskij, and M.~Nagy,
arXiv:1206.3438[hep-ph] (2012).

\bibitem{SBKLN-1207_6937}
Yu.S.~Surovtsev, P.~Byd\v{z}ovsk\'y, R.~Kami\'nski, V.E.~Lyubovitskij, and M.~Nagy,   
arXiv:1207.6937[hep-ph] (2012).

\bibitem{Mark_III}
W.~Lockman (Mark III), Proceedings of the Hadron'89 Conference, ed. F.~Binon {\it et al.} (Editions Fronti{\`e}res, Gif-sur-Yvette,1989) p.109;
A.~Falvard {\it et al.} (DM2), Phys. Rev. {\bf D38} 2706 (1988);
M.~Ablikim {\it et al.} (BES II), Phys. Lett. {\bf B607} 243 (2005).

\bibitem{Mark_II}
G.~Gidal {\it et al.} (Mark~II), Phys. Lett. {\bf B107} 153 (1981);
M.~Oreglia {\it et al.} (Crystal Ball(80)), Phys. Rev. Lett. {\bf 45} 959 (1980).

\bibitem{Argus}
H.~Albrecht {\it et al.} (Argus), Phys. Lett. {\bf B134} 137 (1984);
D.~Gelphman {\it et al.} (Crystal Ball(85)), Phys. Rev. {\bf D32} 2893 (1985);
D.~Besson {\it et al.} (CLEO), Phys. Rev. {\bf D30} 1433 (1984);
V.~Fonseca {\it et al.} (CUSB), Nucl. Phys. {\bf B242} 31 (1984);

\bibitem{Achasov-4q}
N.N.~Achasov, Nucl. Phys. {\bf A675} 279c (2000).

\bibitem{Jaffe-4q}
R.L.~Jaffe, Phys. Rev. {\bf D15} 267, 281 (1977).

\bibitem{MO-02-2}
P.~Minkowski and W.~Ochs, Eur. Phys. J. {\bf C9} 283 (1999); arXiv:hep-ph/0209223 (2002); hep-ph/0209225 (2002).

\bibitem{Ochs10}
W.~Ochs, AIP Conf. Proc. {\bf 1257} 252 (2010); arXiv:1001.4486[hep-ph] (2010).

\bibitem{Bugg1370} D.V.~Bugg, Eur. Phys. J. C {\bf C52} 55 (2007);
arXiv: hep-ex/0706.1341; arXiv: 0710.4452 [hep-ex].

\bibitem{Geng_Oset} L.S.~Geng and E.~Oset,
Phys. Rev. {\bf D79} 074009 (2009); arXiv:0812.1199 [hep-ph].

\bibitem{KMS-nc96}
D.~Krupa, V.A.~Meshcheryakov and Yu.S.~Surovtsev,
Nuovo Cim. {\bf A109} 281 (1996).

\bibitem{Kato-AP65}
M.~Kato, Ann. Phys. {\bf 31} 130 (1965).

\bibitem{Morgan_Penn_93}
D.~Morgan and M.R.~Pennington, Phys. Rev. {\bf D48} 1185, 5422 (1993).

\bibitem{LeCou}
K.J.~Le~Couteur, Proc. R. London, Ser. A {\bf 256} 115 (1960); R.G.~Newton, J. Math. Phys. {\bf 2} 188 (1961).

\bibitem{Zou_Bugg_94}
B.S.~Zou and D.V.~Bugg, Phys. Rev. {\bf D50} 591 (1994).

\bibitem{Guo}
B.~Liu, M.~Buescher, F.-K.~Guo, C.~Hanhart, and U.-G.~Meissner,
Eur. Phys. J. {\bf C63} 93 (2009).

\bibitem{expd_pipi}
J.R.~Batley {\it et~al.}, Eur. Phys. J. {\bf C54} 411 (2008);
B.~Hyams {\it et~al.}, Nucl. Phys. {\bf B64} 134 (1973); {\bf B100} 205 (1975); A.~Zylbersztejn {\it et~al.}, Phys. Lett. {\bf B38} 457 (1972);
P.~Sonderegger and P.~Bonamy, in Proc. 5th Int. Conference on Elementary Particles, Lund, 1969, 372; J.R.~Bensinger {\it et~al.}, Phys. Lett. {\bf B36} 134 (1971); J.P.~Baton {\it et~al.}, Phys. Lett. {\bf B33} 525, 528 (1970);
P.~Baillon {\it et~al.}, Phys. Lett. {\bf B38}, 555 (1972); L.~Rosselet {\it
et~al.}, Phys. Rev. {\bf D15} 574 (1977); A.A.~Kartamyshev {\it et~al.},
Pis'ma Zh. Eksp. Theor. Fiz. {\bf 25} 68 (1977); A.A.~Bel'kov {\it et~al.},
Pis'ma Zh. Eksp. Theor. Fiz. {\bf 29} 652 (1979); S.D.~Protopopescu {\it et~al.}, Phys. Rev. {\bf D7} 1279 (1973); P.~Estabrooks and A.D.~Martin, Nucl. Phys. {\bf B79} 301 (1974).

\bibitem{expd_pipiKK}
W.~Wetzel {\it et al.}, Nucl. Phys. {\bf B115} 208 (1976); V.A.~Polychronakos {\it et~al.}, Phys. Rev. {\bf D19} 1317 (1979); P.~Estabrooks, Phys. Rev. {\bf D19} 2678 (1979); D.~Cohen {\it et~al.}, Phys. Rev. {\bf D22} 2595 (1980); G.~Costa {\it et~al.}, Nucl. Phys. {\bf B175} 402 (1980); A.~Etkin {\it et~al.}, Phys. Rev. {\bf D25} 1786 (1982).

\bibitem{expd_eta_eta}
F.~Binon {\it et al.}, Nuovo Cim. {\bf A78} 313 (1983).

\bibitem{Achasov-Shest}
N.N.~Achasov and G.N.~Shestakov, Phys.~Rev. {\bf D49} 5779 (1994).

\bibitem{SKN-epja}
Yu.S.~Surovtsev, D.~Krupa, and M.~Nagy, Eur.~Phys.~J. {\bf A15} 409 (2002); Czechoslovak Journal of Physics {\bf 56} 807 (2006).

\bibitem{Wei90}
S.~Weinberg, Phys. Rev. Lett. {\bf 65} 1177 (1990).

\bibitem{Nieves-Arriola}
J.~Nieves and E.R.~Arriola, Phys. Rev. {\bf D80} 045023 (2009).

\bibitem{GLSV_13}
T.~Gutsche, V.E.~Lyubovitskij, I.~Schmidt, and A.~Vega,
Phys. Rev. {\bf D87} 056001 (2013); arXiv:1212.5196 [hep-ph] (2012).

\bibitem{Isgur} J. Weinstein and N. Isgur, Phys. Rev. Lett. {\bf 48} 659 (1982); Phys. Rev. {\bf D27} 588 (1983); {\bf D41} 2236 (1990).

\bibitem{Jansen} G. Jansen, B.C. Pearce, K.~Holinde, and J.~Speth, Phys.~Rev. {\bf D52} 2690 (1995).

\bibitem{BGL} T.~Branz, T.~Gutsche, and V.~E.~Lyubovitskij, Eur. Phys. J.
{\bf A37} 303 (2008).

\bibitem{GarciaMKPRE-11}
R.~Garc{\'i}a-Mart{\'i}n, R.~Kami{\'n}ski, R.~Pel{\'a}ez,
and J.~Ruiz~de~Elvira,
Phys. Rev. Lett.  {\bf 107} 072001 (2011).

\bibitem{Amsler95}
C.~Amsler {\it et~al.}, Phys. Lett. {\bf B355} 425 (1995).

\bibitem{Braccini99}
S.~Braccini, Frascati Phys. Series {\bf XV} 53 (1999).

\bibitem{Barate00}
R.~Barate {\it et~al.}, Phys. Lett. {\bf B472} 189 (2000).

\bibitem{Anisovich97} V.V.~Anisovich {\it et~al.}, Nucl. Phys. A (Proc Suppl.)
{\bf 56} 270 (1997).

\bibitem{Ellis-Lanik} J. Ellis and J. L\'anik, Phys. Lett. {\bf B150}
289 (1985).

\bibitem{Parganlija-glbl} S.~Janowski, D.~Parganlija, F.~Giacosa, and D.~H.~Rischke, Phys. Rev. {\bf D84} 054007 (2011).

\bibitem{GGLF} F.Giacosa, T.~Gutsche, V.~E.~Lyubovitskij, and
A.~Faessler, Phys. Rev. {\bf D72}, 094006 (2005).

\bibitem{SBGL-PPN10} Yu.S.~Surovtsev, T.~Branz, T.~Gutsche, and V.E.~Lyubovitskij, Physics of Particles and Nuclei {\bf 41} 990 (2010).


\end{thebibliography}
\end{document}